\documentclass[10pt]{article}
\usepackage[T1]{fontenc}
\usepackage[latin9]{inputenc}
\usepackage[verbose,tmargin=0.75in,bmargin=0.75in,lmargin=1in,rmargin=0.75in]{geometry}
\usepackage{color}
\usepackage{babel}
\usepackage{multirow}
\usepackage{amsmath}
\usepackage{amssymb}
\usepackage{graphicx}
\usepackage{esint}
\usepackage[unicode=true, bookmarks=false,
breaklinks=false,pdfborder={0 0 1},backref=section,colorlinks=false]{hyperref}

\makeatletter

\usepackage{cite}
\usepackage{setspace}
\usepackage{abstract}
\usepackage{appendix}
\usepackage{diagbox}
\usepackage{multirow}
\usepackage{enumitem}
\usepackage[affil-it]{authblk}

\usepackage{calc}
\usepackage{babel}

\newcommand{\be}{\begin{equation}}
\newcommand{\ee}{\end{equation}}

\newcommand{\mathDf}{\delta\!f}
\newcommand{\HoptA}{h_{\rm opt,av}}
\newcommand{\QminA}{Q_{\rm min,av}}
\newcommand{\Hopt}{h_{\rm opt}}
\newcommand{\Qmin}{Q_{\rm min}}

\title{\vspace{-1cm} Noise and error analysis and optimization in particle-based kinetic plasma simulations}
\author[1]{E.~G. Evstatiev\footnote{Corresponding author: \href{mailto:egevsta@sandia.gov}{egevsta@sandia.gov}}}
\author[2]{J.~M. Finn}
\author[3]{B.~A. Shadwick}
\author[4]{N. Hengartner}
\affil[1]{Sandia National Laboratory, Albuquerque, NM}
\affil[2]{Tibbar Plasma Technologies, Los Alamos, NM}
\affil[3]{University of Nebraska-Lincoln, Lincoln, NE}
\affil[4]{Los Alamos National Laboratory, Los Alamos, NM}

\date{16 August 2020}

\begin{document}

\maketitle
\begin{abstract}
In this paper we analyze the noise in macro-particle methods used in plasma
physics and fluid dynamics, leading to approaches for minimizing the
total error, focusing on electrostatic models in one dimension.
We begin by describing kernel density estimation for continuous values
of the spatial variable $x$, expressing the kernel in a form in which
its shape and width are represented separately. The covariance
matrix $C(x,y)$ of the noise in the density is computed, first for
uniform true density. The band width of the covariance matrix
is related to the width of the kernel. A feature that stands out is
the presence of constant negative terms in the elements of the covariance
matrix both on and off-diagonal.
These negative correlations are related to the fact that the
total number of particles is fixed at each time step; they also lead
to the property $\int\!\, C(x,y)\,dy=0$.
We investigate the effect of these
negative correlations on the electric field computed by Gauss's law,
finding that the noise in the electric field is related to a process
called the \emph{Ornstein-Uhlenbeck bridge}, leading to a covariance
matrix of the electric field with variance significantly reduced relative
to that of a Brownian process.

For non-constant density, $\rho(x)$, still with continuous $x$, we
analyze the total error in the density estimation and discuss it in
terms of \emph{bias-variance optimization} (BVO).
For some characteristic length $l$, determined by the density and its second derivative,
and kernel width $h$, having too few particles within $h$
leads to too much variance; for $h$ that is large
relative to $l$, there is too much smoothing of the density.
The optimum between these two limits is found by BVO. For kernels of the same
width, it is shown that this optimum (minimum) is weakly sensitive
to the kernel shape.

We repeat the analysis for $x$ discretized on a grid. In this case
the charge deposition rule is determined by a \emph{particle shape}.
An important property to be respected in the discrete system is the
exact preservation of total charge on the grid; this property
is necessary to ensure that the electric field is equal at both ends,
consistent with periodic boundary conditions. 
We find that if the particle shapes satisfy a sum rule, the particle
charge deposited on the grid is conserved exactly.
Further, if the particle shape is expressed
as the convolution of a kernel with another kernel that satisfies
the sum rule, then the particle shape obeys the sum rule. This property
holds for kernels of arbitrary width, including widths that
are not integer multiples of the grid spacing.

We show results relaxing the approximations used to do BVO optimization analytically,
by doing numerical computations of the total error as a function of
the kernel width, on a grid in $x$. The
comparison between numerical and analytical results shows good agreement
over a range of particle shapes.

We discuss the practical implications of our results, including
the criteria for design and implementation of computationally efficient particles
that take advantage of the developed theory.

\end{abstract}

\section{Introduction}
\label{sec:introduction}

The particle-in-cell (PIC) method has been an indispensable tool of
numerical modelers in fluid dynamics and kinetic plasma physics for
several decades now \cite{harlow:1964,hockney:1826,langdon_birdsall:2115,
  lewis:1970:136,Brackbill-Ruppel,dawson_particle_1983,Hockney88,birdsall_plasma_2004}
and the variety of kinetic plasma problems to which it has been applied
keeps increasing. The success of this method has spurred more recent
developments \cite{chen_chacon:2011,markidis_lapenta:2011,
  squire_geometric_2012,evstatiev_variational_2013,shadwick_variational_2014,stamm_variational_2014},
with emphasis on geometrical aspects; for example, advantage has been
taken of the Hamiltonian nature of the Vlasov-Maxwell system
\cite{morrison:1980:383,Weinstein-Morrison81}.
To ensure conservation properties such as momentum, energy, charge,
etc., some formulations \cite{lewis:1970:136,squire_geometric_2012,evstatiev_variational_2013,kraus_gempic_2017}
rely on a variational method, related to the Hamiltonian prescription,
while others devise specific spatial and temporal discretizations for
providing conservation properties \cite{chen_chacon:2011,markidis_lapenta:2011,kraus_gempic_2017}.

The first particle methods applied to plasma simulations \cite{hockney:1826}
showed significant effects of noise due to a few factors:
first, the deposition of a particle's charge applied to only one grid
node, an approach called the nearest grid point (NGP) method; second,
very few particles were used, due to the limited computational power
in the early 1970s; third, guidelines for noise minimization were
quite limited.

The introduction of finite size computational particles by Birdsall
and Langdon \cite{langdon_birdsall:2115} significantly elevated the usefulness
of the particle method. Indeed, a hierarchy of shapes,
varying in size and smoothness, were proposed to address issues of
noise \cite{birdsall_clouds_1969,langdon_kinetic_1979} as well as frequency aliasing
\cite{birdsall_plasma_2004}.

The recognition that the level of particle noise scales as $N_{p}^{-1/2}$,
where $N_{p}$ is the number of computational particles, together
with ever-increasing demands for accuracy and fidelity of simulations
has again put the issue of noise in particle methods in the spotlight.
Advancement has come with the emergence of hybrid kinetic-fluid methods
and the $\mathDf$ method
\cite{Kotschenreuther-1988,Kotschenreuther-deltaf,parker_fully_1993,hu_generalized_1994,brunner_collisional_1999}.
In spite of this progress, noise is still a limiting factor
in particle codes: in $\mathDf$ and hybrid methods particles are used to describe
only a subset of the distribution function, however, noise is still an important factor
for the particle part of the computations as well as for the fluid-particle
coupling (for time-evolving fluids).
In full kinetic treatments as well
in hybrid and $\mathDf$ methods, noise 
in the density is an especially serious problem for quasineutral plasmas, 
in which the local net charge density is small.

The present work presents a new analysis of the statistics of noise
in particle-based methods. Specifically, we analyze the error in
the estimation of the particle density in terms of a finite number
of computational particles of finite size, a special case of
\emph{kernel density estimation}, with a focus on the \emph{bias-variance trade-off}
\cite{bishop_pattern_2006,Sammut_encyclopedia_2017}.
We also analyze the error in the electric field computed from the
charge density, showing that certain negative correlations in the
density noise lead to properties of the electric field related to
the \emph{Ornstein-Uhlenbeck bridge} \cite{Brownian_2nd_order2,Corlay_arxiv_2013,Mazzoloa_2017},
a generalization of the \emph{Brownian bridge} \cite{revuz_continuous_1999},
a Brownian process with boundary conditions at each end. We concentrate
on a 1D (one-dimensional) electrostatic (ES) formulation with periodic
boundary conditions, with overall charge neutrality and immobile ions, leaving
generalizations such as to higher dimensions and electromagnetic models
for future work.

In Sec.~\ref{sec:density_estimation} we establish the framework
used in estimating the electron density and its noise properties
in a $1$D electrostatic Vlasov-Poisson system. In this system, the
only source of noise is the estimated charge density. (Electromagnetic
models also involve noise in the estimated current density.) These issues related
to density estimation are introduced with a continuous,
i.e. non-discretized, spatial variable $x$. We also
discuss the various kernels that can be used, show how the kernel
width and its shape (smoothness) enter, and summarize some properties
of these kernels that relate to discretization and that will enter in later sections.

In Sec.~\ref{sec:stat_uniform}, and in the next section, we continue to restrict
our attention to continuous $x$. We introduce the covariance
matrix for the noise in the density, focusing in this section on a
system in which the ``true'' density is uniform. We discuss the
origin of certain negative terms in the covariance matrix.
(These negative off-diagonal terms represent negative correlations.)
We show that in computing the electric field by Gauss's law, these
negative correlations and the boundary conditions on the electric
field lead to properties associated with the Ornstein-Uhlenbeck bridge. We characterize
noise in the computed electric field in terms of its covariance matrix.
We illustrate with a kernel involving a delta function. Issues associated
with the Brownian bridge are discussed in more depth in Appendix~\ref{app:brownian_bridge}
and issues related to relaxing the delta function restriction to give a nonzero
kernel width are
discussed in Appendix~\ref{app:electric_field_covariance}.

In Section \ref{sec:stat_nonuniform} we generalize to non-uniform
density and discuss the
application of bias-variance optimization \cite{bishop_pattern_2006}
to find the optimal kernel width. This optimization in the presence
of non-uniform density minimizes the total error in the density estimated
with a kernel of a specific shape and width. This error consists of
a \emph{variance} term (noise) caused by the finite number of particles
and a \emph{bias} term, a smoothing of the density that occurs because
of the finite kernel width. For the remainder of this paper we refer to
\emph{noise} as the error due to having a finite number of particles
and the more general term \emph{error} as including the bias.
Issues relating to the scaling of the kernel width are discussed in
Appendix \ref{app:kernel_scaling}.

In Section \ref{sec:grid_discretization} we discuss the density and
electric field on a discrete grid, where a particle shape for the
charge deposition enters. We discuss the importance of a sum rule;
obeying this sum
rule is a sufficient condition for the net charge on the grid to be
exactly zero when the ion charge is subtracted. This requirement assures
that the electric field at the endpoints are equal. We also show that
for a general kernel, the sum rule is obeyed if the particle shape
is a convolution of the kernel with another kernel that already satisfies
the sum rule. We discuss the covariance matrix of the noise on the
discrete grid for various particle shapes.

In Sec.~\ref{sec:numerical} we compute the total error (bias plus
variance) numerically for various shapes and
compare with the analytic theory of sections~\ref{sec:stat_uniform}
and \ref{sec:stat_nonuniform}.

In Sec.~\ref{sec:conclusions} we summarize and discuss our results.

\section{Kernel density estimation by a finite number of particles}
\label{sec:density_estimation}

Particle methods are hybrid Lagrangian-Eulerian in nature: computational
macro-particles are allowed to move with continuous positions and
velocities, while charge densities and fields are resolved on a fixed computational
grid. The connection between particles and grid is via a particle
shape, which specifies a charge deposition rule. It has been traditional
\cite{birdsall_plasma_2004} to use particle sizes that are integer
number of computational cells wide, although such restriction is not
necessary; a related issue, which we also discuss in Sec.~\ref{sec:sum_rule},
is that the width of a particle shape and its degree of smoothness
need not be related. The latter point has been emphasized in
Ref.~\cite{evstatiev_variational_2013}.
Furthermore, grid size is many times determined subjectively
by the modeler according to a desired resolution, accuracy, the particular
physics problem under consideration, etc. This resolution may be increased
a few times to determine convergence of the numerical results,
while also increasing the number of particles; a typical quantity
that is kept constant is the average number of particles per cell.
Of course, every time the grid resolution and particle number are
increased, the demand for computational resources increases and for
large problems this strategy quickly becomes prohibitive. The analysis
in this paper aims to provide a systematic way of minimizing noise
and error in the charge density by selecting
optimal size and number of particles and, as a consequence, to minimize
the computational resources required to achieve a given accuracy.

Our discussion will focus on 1D electrostatic models, in periodic geometry.
As will be shown in the following, the charge density and its gradients
are essential for the analysis of noise and error. Therefore, we consider
working with quantities that are periodic functions of $x$ on the real interval
$[0,1]$, but over all real values of
the particle velocity $v$. This choice is advantageous for the presentation
of the ideas, postponing grid discretization to later sections.

We introduce a representation of the electron distribution function in phase
space $(x,v)\in[0,1]\times(-\infty,\infty)$ in terms of $N_{p}$
number of finite-size computational particles, 
\begin{equation}
  f_{e}(x,v,t)=\sum_{\mu=1}^{N_{p}}q_{\mu}K(x-\xi_{\mu})\delta(v-\dot{\xi}_{\mu})\,,
  \label{eq:dist_function}
\end{equation}
where $f_{e}$ is the \emph{estimated} phase space distribution, $q_{\mu}$
is the computational particle charge, 
$\xi_{\mu}$ is the computational particle position, and $\dot{\xi}_{\mu}$ is its
velocity. We use $q_\mu>0$ in Eq.~\eqref{eq:dist_function} and throughout the presentation
and the negative sign of the electron charge is added explicitly
in places where it is used, e.g., in Gauss's law (so strictly speaking $q_{\mu}$
is the weight and the macro-particle charge is $-q_{\mu}$).
The general form of the kernel is $K(x,\xi)$, however, due to the
periodic boundary conditions it assumes the translationally invariant form $K(x-\xi)$
(see below).
The subscript ``\textit{e}'' in Eq.~\eqref{eq:dist_function} and everywhere throughout
the paper stands for ``estimated.''
Also, for the rest of the paper we use the term \emph{particle} in place of \emph{computational
particle} or \emph{macro-particle.} This particle is usually comprised of many physical particles.

By integrating over velocity space, we obtain the
\emph{estimated density} of the electrons
at any spatial point $x$ in terms of the positions of all of the
particles: 
\begin{equation}
\rho_{e}(x)=\sum_{\mu=1}^{N_{p}}q_{\mu}K(x-\xi_{\mu})\,.\label{eq:rho_estimated}
\end{equation}
We take the special case, in which all the $q_{\mu}$ are equal. With
periodic boundary conditions on $[0,1]$, no particles are gained
or lost, so $\int_{0}^{1}\rho_{e}(x)\,dx$ is conserved. We normalize
to $\int_{0}^{1}\rho_{e}(x)dx=1$ and assume immobile
ions with uniform, fixed density $\rho^{(i)}(x)$. Overall neutrality is assumed,
i.e.~$\int_{0}^{1}\rho^{(i)}(x)dx=1$ as well.

The form in Eq.~(\ref{eq:rho_estimated}) is the usual form of kernel
density estimation, used in statistics and machine learning\cite{bishop_pattern_2006}.
The kernel $K(x)$ is usually assumed to satisfy the following conditions,
which do not present practical limitations: 
\begin{align}
 & \mbox{\textbullet\,\, Normalized to unity,}\nonumber \\
 & \mbox{\hskip20pt}\int_{0}^{1}\!\,K(x)\,dx=1\,;\label{eq:S_normalization-1}\\
 & \mbox{\textbullet\,\, Symmetric, }K(x)=K(-x),\quad x\in[0,1];\label{eq:S_symmetry-1}\\
 & \mbox{\textbullet\,\, Translationally invariant, }K(x,\xi)=K(x-\xi),\quad x,\xi\in [0,1];\label{eq:S_translation-1}\\
 & \mbox{\textbullet\,\, Nonnegative, }K(x)\ge0,\quad x\in[0,1];\label{eq:S_positive-1}\\
 & \mbox{\textbullet\,\, Has compact support.}\label{eq:S_finite-1}
\end{align}
Condition \eqref{eq:S_normalization-1} ensures the density normalization discussed above
while conditions \eqref{eq:S_symmetry-1}--\eqref{eq:S_finite-1}
are chosen out of convenience but are not essential for the theory development.
The normalization condition on the kernel and the condition $\int\rho_{e}(x)dx=1$
imply $\sum_{\mu=1}^{N_{p}}q_{\mu}=1$ while the assumed equal and constant particle
charges lead to $q_{\mu}=1/N_{p}$.

At this stage, there is no grid, so the kernel
width\footnote{We will call the measure of the support the \emph{width} of the kernel.}
is not related to a grid spacing. We, in fact, express a kernel of
width $h$ as 
\begin{equation}
  K(x)=\frac{1}{h}K_{f}\left(\frac{x}{h}\right),
  \label{eq:K_definition}
\end{equation}
where $K_{f}$ is the fundamental kernel with support $[-1/2,1/2]$.
Thus, $K_{f}$ contains all the information on the particle shape, including
its smoothness, while its width is independently set by  $h$.
(To be specific, we assume that $K_{f}$ is defined on the real line, and
after scaling to form $K(x)$, it is extended to be periodic with
period unity.)

\begin{figure}[!t]
\begin{center}
\includegraphics[width=0.7\textwidth]{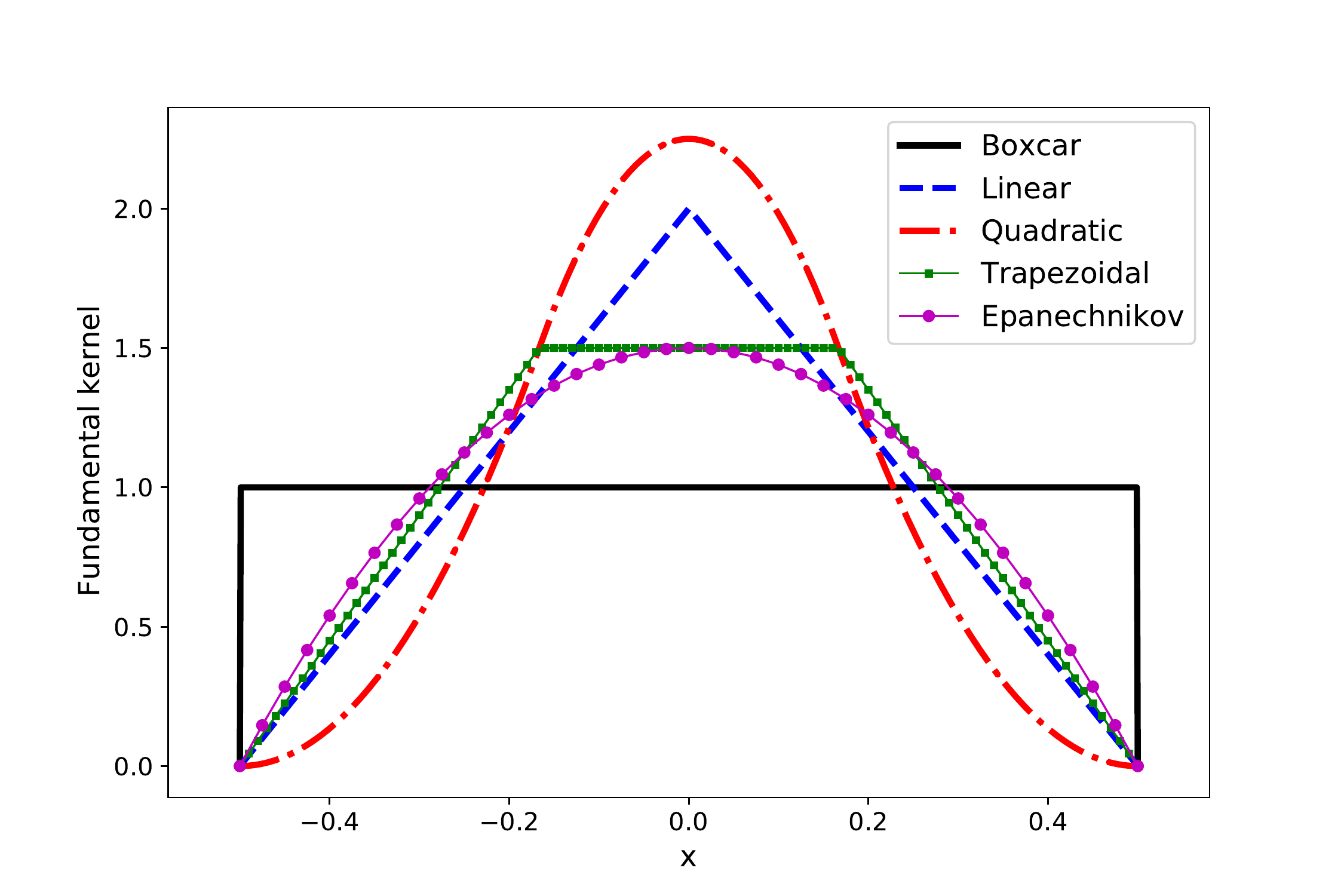}
\end{center}
\caption{Illustration of the fundamental kernel examples from Table~\ref{table:fundamental_kernels},
  normalized to unity. The important issues of bias-variance optimization
  and the sum rule in Eq.~\eqref{eq:S_sum_rule} are discussed in the text.}
\label{fig:fundamental_kernels}
\end{figure}

Examples of fundamental kernels are given in
Table~\ref{table:fundamental_kernels} and illustrated in Fig.~\ref{fig:fundamental_kernels}.
The boxcar, linear and quadratic kernels are the convolutional particle shapes
of Ref.~\cite{Hockney88,birdsall_plasma_2004}, 
scaled to the unit interval; the trapezoidal kernel 
is discussed in the following sections.
Another important kernel is the Epanechnikov kernel \cite{epanechnikov_non-parametric_1969}.
Sec.~\ref{sec:optimal_kernel} discusses the BVO process,
optimal kernel width, etc., where the shape factors
$\int\! K_{f}(x)^{2}\,dx$ and $\int\! x^2 K_{f}(x)\,dx$,
which are of order unity, play a prominent role.

In sections \ref{sec:grid_discretization} and \ref{sec:numerical} we will apply these results
involving the kernel $K(x)$ in the presence of a uniform grid 
$x_i$, $i=1,2,\ldots,N_g$ with
spacing\footnote{We restrict our attention to uniform grids strictly for convenience.}
$\Delta=1/N_g$. There we will construct a \emph{particle shape} $S(x)$, which
satisfies conditions \eqref{eq:S_normalization-1}-\eqref{eq:S_finite-1} for
a kernel and also satisfies a \emph{sum rule}
\begin{equation}
\sum_{i=1}^{N_{g}}\Delta\,S(x_{i}-\xi)=1\,\label{eq:S_sum_rule}
\end{equation}
for an arbitrary particle position $\xi$, the discrete analog of
the normalization condition in Eq.~(\ref{eq:S_normalization-1}).
We will show that a sufficient condition 
for $S(x)$ to satisfy the sum rule is that it be a convolution
of a kernel of arbitrary width $\delta$ with either
another particle shape or a finite element of width $i\Delta,\,i=1,2,\dots$;
thereby, $S(x)$ can have an arbitrary width $h=i\Delta+\delta$.

As a last comment in this section, we note that not every kernel when scaled to a grid
satisfies the sum rule; among our examples, the Epanechnikov
kernel scaled to the grid spacing (width $h=i\Delta+\delta$) 
does not obey the sum rule (\ref{eq:S_sum_rule}) for any $\delta$ and $i=1,2,\dots$.
In contrast, the boxcar, linear, quadratic, and trapezoidal kernels, when scaled to
$h =\Delta,\, 2\Delta,\, 3\Delta$, $3\Delta$, correspondingly [cf.~Eq.~\eqref{eq:K_definition}],
do satisfy the sum rule. We shall discuss these issues further in Sections \ref{sec:grid_discretization}
and \ref{sec:numerical}. At the same time, since all particle shapes
satisfy conditions \eqref{eq:S_normalization-1}--\eqref{eq:S_finite-1}, any particle shape
may be used as a kernel in the density estimation expression \eqref{eq:rho_estimated}.

\begin{table}
  \footnotesize
  \begin{center}
    \begin{tabular}{l|l}
      \hline \rule{0pt}{3ex} 
    Kernel & Definition   \\
    \hline\hline \rule{0pt}{5ex} 
    Boxcar (top-hat) & $ K_{fB}(x) =
                       \left\{
                       \begin{tabular}{ll}
                         \ensuremath{1},    \ensuremath{|x|\le\frac{1}{2}}\\[1.25ex]
                         \ensuremath{0}    otherwise\,. 
                       \end{tabular} \right. $ \\[1.25em]
    \hline \rule{0pt}{5ex} 
    Linear (tent) & $ K_{fL}(x) =
                    \left\{
                    \begin{tabular}{ll}
                      \ensuremath{2\left(1-2|x|\right)},    \ensuremath{|x|\le\frac{1}{2}}\\[1.25ex]
                      \ensuremath{0}    otherwise\,. 
                    \end{tabular}\right. $ \\[1.25em]
    \hline \rule{0pt}{7ex}
    Quadratic &  $ K_{fQ}(x) =
                9\left\{
                \begin{tabular}{ll}
                  \ensuremath{\frac{1}{4}-3x^{2}}, \ensuremath{\left|x\right|\le1/6}\\[1.25ex]
                  \ensuremath{\frac{3}{2}\left(\frac{1}{2}-|x|\right)^{2}}, \ensuremath{1/6\le\left|x\right|\le1/2} \\[1.25ex]
                  \ensuremath{0}    otherwise\,. 
                \end{tabular}\right. $ \\[2em]
    \hline \rule{0pt}{7ex}
    Trapezoidal & $ K_{fT}(x) =
                  \frac{3}{2}\left\{
                  \begin{tabular}{ll}
                    1,   \ensuremath{|x|\le1/6}\\[1.25ex]
                    \ensuremath{3\left(\frac{1}{2}-|x|\right)}, \ensuremath{1/6\le|x|\le1/2} \\[1.25ex]
                    \ensuremath{0}   otherwise\,. 
                  \end{tabular}\right. $ \\[2em]
    \hline \rule{0pt}{5ex}
    Epanechnikov & $ K_{fE}(x) =
                   \left\{
                   \begin{tabular}{ll}
                     \ensuremath{\frac{3}{2}\left(1-4x^{2}\right)},   \ensuremath{|x|\le\frac{1}{2}}\\[1.25ex]
                     \ensuremath{0}   otherwise\,, 
                   \end{tabular}\right. $ \\[1.25em]
    \hline\hline
  \end{tabular}
\end{center}
\caption{Examples of fundamental kernels, with width unity and normalized to have $\int K(x)dx=1$.}
\label{table:fundamental_kernels}
\end{table}

\section{Statistical analysis in uniform density}
\label{sec:stat_uniform}

Based on the kernel representation from Sec.~\ref{sec:density_estimation},
we analyze in this section what is typically known as ``noise''
in the PIC method. We note that---as will become clear in the next
section---noise is only one part of the total error that we make
when estimating the ``true'' density (or field) with the
help of a finite number of particles, the other part being the \emph{bias}
error. When estimating the error in a uniform density, as we do in
this section, only the noise part appears, the bias part being zero
for this case. The focus is on the covariance matrix between the noise
in the density at different spatial points. Based on the density 
covariance matrix, we consider
the covariance matrix between the noise in the electric field at different
spatial points.

\subsection{Statistical analysis of the estimated density}
\label{sec:correlations_uniform}

In this section we introduce the mathematical method of the analyses
of noise, and later that of error, while also deriving
the uniform density correlations with
the important negative contributions resulting from 
the fixed number of particles in a numerical simulation.
Let us denote the \emph{true} electron density by $\rho(x)$.
Because of our choice of normalization, the true electron density (or true density)
satisfies all the properties of a
\emph{probability density function}, and the normalization
condition \eqref{eq:S_normalization-1} guarantees that $\rho_e(x)$ does also.
As discussed above, at this point in our analysis we do not consider
grid discretization and hence we do not relate the kernel width $h$
to the grid spacing $\Delta$. Using the continuous spatial variable $x$,
the average of any quantity $f(x)$ is calculated as
\begin{equation}
  \left<f\right>=\int_{0}^{1}\!\!\,f(\xi)\rho(\xi)\,d\xi
  \label{eq:f_x_ave}
\end{equation}
or
\begin{equation}
  \left<f\right> = \int_0^1\!\!f(\xi,\eta)\rho(\xi)\rho(\eta)\,d\xi d\eta
  \label{eq:f_xy_ave}
\end{equation}
for a function of two random variables, etc.
In this way, we can calculate the \emph{expected} (statistical expectation)
value of the estimated density over the true density $\rho(x)$ as
\begin{align}
  \langle\rho_{e}(x)\rangle=& \left(\sum_{\mu}q_{\mu}\right)\int_0^1 K(x-\xi)\rho(\xi)d\xi \nonumber\\
  =& \int_{x-h/2}^{x+h/2} K_{f}\left(\frac{x-\xi}{h}\right)\rho(\xi)\frac{d\xi}{h} 
  = \int_{-1/2}^{1/2} K_{f}(\eta)\rho(x+h\eta)d\eta,
  \label{eq:<rhoe>}
\end{align}
where $\sum_\mu q_\mu=1$ has been used, the symmetry of $K$ has been used,
and we have defined $\eta = (\xi - x)/h$. Expanding for small $h$, we find
\begin{align}
  \langle\rho_{e}(x)\rangle
  &= \rho(x)\int_{-1/2}^{1/2} K_{f}(\eta)d\eta
     +\frac{h^{2}}{2}\rho''(x)\int_{-1/2}^{1/2} K_{f}(\eta)\eta^{2}d\eta+\cdots \nonumber \\
  &= \rho(x) + \frac{h^{2}}{2}\rho''(x)\int_{-1/2}^{1/2} K_{f}(\eta)\eta^{2}d\eta+\cdots \,.
  \label{eq:<rhoe>-2}
\end{align}
where the normalization $\int_{-1/2}^{1/2}K_f(\eta) d \eta=1$ condition has been used and
the symmetry of $K_{f}$ has again been used to conclude $\int_{-1/2}^{1/2} K_{f}(\eta)\eta d\eta=0$.
(Henceforth, we omit integral limits to improve readability.)
We defer the issues of a non-uniform density $\rho(x)$ to a later section.

For the uniform density case considered in this section, $\rho(x)=1$
(hence $\rho''(x)=0$), and we obtain 
\[
\langle\rho_{e}(x)\rangle=\rho(x)=1.
\]
Now defining the fluctuations $\tilde{\rho}_e(x)$ by
$\rho_{e}(x)=\langle\rho_{e}(x)\rangle+\tilde{\rho}_{e}(x)=1+\tilde{\rho}_{e}(x)$
(note that $\left<\tilde{\rho}_e(x)\right>=0$),
we can write the \emph{variance} as $V(x)=\langle\tilde{\rho}_{e}(x)^{2}\rangle=\langle\rho_{e}(x)^{2}\rangle-1$, or 
\[
V(x)=V_{d}(x)+V_{o}(x)-1
\]
where $V_{d}(x)$ and $V_{o}(x)$ denote the diagonal ($\mu=\nu$)
and off-diagonal ($\mu\ne\nu$) terms in $V(x)$, as defined below.
We find 
\begin{equation}
  V_{d}(x)=\sum_{\mu}q_{\mu}^{2}\langle K(x-\xi_{\mu})^{2}\rangle
  =\frac{1}{N_{p}}\int K(x-\xi)^{2}d\xi
  =\frac{1}{N_{p}}\int K(\xi)^{2}d\xi
  \label{eq:diag_V}
\end{equation}
and 
\begin{align}
  V_{o}(x)=&\, \sum_{\mu\ne\nu}q_{\mu}q_{\nu}\langle K(x-\xi_{\mu})K(x-\xi_{\nu})\rangle \nonumber \\
  =&\, \frac{N_{p}(N_{p}-1)}{N_{p}^{2}}\int K(x-\xi)K(x-\eta)d\xi d\eta \nonumber \\
  =&\, \frac{N_{p}(N_{p}-1)}{N_{p}^{2}}\left(\int K(\xi)d\xi\right)^{2}, \nonumber \\
  =&\, 1-\frac{1}{N_{p}}.
  \label{eq:offdiag_V}
\end{align}
At this point we notice the general scaling
$V_{d}=(1/N_{p}h)\int K_{f}(\eta)^{2}d\eta\sim1/N_{p}h$, which follows
by using Eq.~(\ref{eq:K_definition}).
The quantity $N_p h$ is the expected number of particles over the width of the kernel.
These results lead to 
\begin{equation}
V(x)=V_{d}(x)-1/N_{p}.\label{eq:Final-variance}
\end{equation}

To relate the estimated density at arbitrary spatial points, $x$ and $y$,
we must compute the \emph{covariance matrix}
\begin{equation}
  C(x,y)=\langle\tilde{\rho}_{e}(x)\tilde{\rho}_{e}(y)\rangle=\langle\rho_{e}(x)\rho_{e}(y)\rangle-1\,.
  \label{eq:Cxy_definition}
\end{equation}
We again use a decomposition into diagonal and off-diagonal terms:
\begin{equation}
  C(x,y)=C_{d}(x,y)+C_{o}(x,y)-1\,.
  \label{eq:Cxy_general}
\end{equation}
The two contributions are
\begin{equation}
  C_{d}(x,y)=C_d(x-y)=\frac{1}{N_{p}}\int K(x-\xi)K(y-\xi)d\xi=\frac{1}{N_{p}}\hat{K}(x-y),
  \label{eq:C_d_xy}
\end{equation}
where $\hat{K}$ is the convolution of $K$ with itself (a legitimate kernel according 
to Eqs.~\eqref{eq:S_normalization-1} - \eqref{eq:S_finite-1}), and
\begin{align}
C_{o}(x,y)&=  \frac{N_{p}(N_{p}-1)}{N_{p}^{2}}\int K(x-\xi)K(y-\eta)d\xi d\eta\nonumber\\
& =  \frac{N_{p}(N_{p}-1)}{N_{p}^{2}}\left(\int K(\xi)d\xi\right)^{2}\nonumber\\
& =  1-\frac{1}{N_{p}} \,.
      \label{eq:C_o_xy}
\end{align}
Putting \eqref{eq:Cxy_general}, \eqref{eq:C_d_xy}, and \eqref{eq:C_o_xy} together, we find 
\begin{equation}
  C(x,y)= \frac{1}{N_p}\left[ \hat{K}(x-y)- 1\right].
  \label{eq:continuum-covariance}
\end{equation}
In the special case $K(x)=\delta(x)$ we have $\hat{K}(x-y)=\delta(x-y)$
and from Eq.~\eqref{eq:continuum-covariance} we obtain 
\begin{equation}
  C(x-y)=\frac{1}{N_{p}}\left[\delta(x-y)-1\right].
  \label{eq:delta-khat}
\end{equation}

Notice the translational invariance form $C(x-y)$ and the presence of constant negative
contributions $-1/N_{p}$ in the expressions for the variance ($x=y$) and the off-diagonal
(correlation) terms ($x\ne y$). In particular, we have
\begin{equation}
  \int\!\!\, C(x,y)\,dy = \int\!\!\, C(x-y)\,dy =0 \,.
  \label{eq:intCdy=00003D00003D0}
\end{equation}
Indeed, $\int C_{d}(x-y)dy=(1/N_{p})\int \!\hat{K}(x-y)\,dy=1/N_{p}$ since
$\int\!\!\,\hat{K}(x-y)\,dx = 1$.
We emphasize that the property in Eq.~\eqref{eq:intCdy=00003D00003D0}
is general, i.e., for any kernel. An alternative proof is given as follows.
Recalling that $\int\! \rho_e(y)\,dy = 1$, it follows
that $\int\! \tilde{\rho}_e(y)\,dy = \int\! (\rho_e(y)-1)\,dy = 0$.
Then from the definition of correlations \eqref{eq:Cxy_definition} we obtain
\[ \int\!\! C(x,y)\,dy = \left<\tilde{\rho}_e(x)\int\!\!\tilde{\rho}_e(y)\right>\, dy = 0\,.   \]

The result Eq.~\eqref{eq:intCdy=00003D00003D0} implies that the function
defined by $u(x)=1$ is in the null space of the
covariance matrix, i.e., is the eigenfunction with zero eigenvalue.

The significance of the negative correlations is further discussed in the next section.

\subsection{Statistical analysis of the electric field}
\label{sec:Electric-field-computation}

In particle codes, noise in the density leads to noise in the electric
field, which in turn affects particle orbits. In this section we quantify
the effect of density noise on the electric field. The quantification
of errors in particle orbits due to errors in the electric field, leading
in turn to density errors, i.e., ``closing
the loop,'' will be the subject of future work.

In the electrostatic model the electric field is computed from Gauss's
law (we use the dimensionless form), 
\begin{equation}
\frac{dE}{dx}=\rho^{(i)}-\rho_{e}=\rho_{q},\label{eq:Gauss}
\end{equation}
where $\rho_{q}$ is the estimated density,
$\rho_{e}$ is the estimated electron density, and again, $\rho^{(i)}=1$
is the fixed background ion density. Because of the assumption that the
net charge is exactly zero, $\int_{0}^{1}\rho_{q}(x)dx=0$,
we have $E(0)=E(1)$, consistent with the assumed periodic boundary conditions.
We also specify that there is no applied potential across the system,
so that 
\begin{equation}
\int_{0}^{1}E(z)dz=0.\label{eq:ZeroPotential}
\end{equation}

To incorporate condition \eqref{eq:ZeroPotential} in the solution
of \eqref{eq:Gauss}, we start with the general expression 
\begin{equation}
\hat{E}(x)=\int_{x_{0}}^{x}\!\!\rho_{q}(z)\,dz\,,\label{eq:basic_Ex}
\end{equation}
where $\hat{E}$ satisfies Eq.~\eqref{eq:Gauss} and $x_0$ is an arbitrary
initial point of integration. We calculate the
integral of $\hat{E}(x)$ over the periodic domain $[0,1]$: 
\begin{align*}
  R_{0}= & \int_{0}^{1}\!\!\hat{E}(x)\,dx=\int_{0}^{1}\!\!dx\,\int_{x_{0}}^{x}\!\!\rho_{q}(z)\,dz
           =\left.x\,\int_{x_{0}}^{x}\!\!\rho_{q}(z)\,dz\right|_{x=0}^{x=1}-\int_{0}^{1}\!\!x\rho_{q}(x)\,dx\\
  = & \int_{x_{0}}^{1}\!\!\rho_{q}(z)\,dz-\int_{0}^{1}\!\!z\rho_{q}(z)\,dz
      =\int_{0}^{1}\!\!\rho_{q}(z)\,dz-\int_{0}^{x_{0}}\!\!\rho_{q}(z)\,dz-\int_{0}^{1}\!\!z\,\rho_{q}(z)\,dz\\
= & -\int_{0}^{x_{0}}\!\!\rho_{q}(z)\,dz-\int_{0}^{1}\!\!x\,\rho_{q}(x)\,dx,
\end{align*}
where in the last line we have used $\int_{0}^{1}\!\!\rho_{q}(z)\,dz=0$.
The quantity $R_{0}$ depends on $x_{0}$. The quantity
$R_{0}$ needs to be subtracted from $\hat{E}(x)$ in order to obtain
an expression $E(x)$ satisfying Eq.~\eqref{eq:ZeroPotential}:
\begin{align}
E(x)= & \hat{E}(x)-R_{0}\nonumber \\
  = & {\int_{x_{0}}^{x}\!\!\rho_{q}(z)\,dz}+{\int_{0}^{x_{0}}\!\!\rho_{q}(z)\,dz}
      +\int_{0}^{1}\!\!x\,\rho_{q}(x)\,dx\nonumber \\
= & \int_{0}^{x}\!\!\rho_{q}(z)\,dz+\int_{0}^{1}\!\!x\rho_{q}(x)\,dx 
\equiv \,\,E_{1}(x)+E_{0}\,. \label{eq:E_for_arbitrary_x0}
\end{align}
Expression \eqref{eq:E_for_arbitrary_x0} indicates the unsurprising
fact that in a periodic system the initial point of integration $x_{0}$
can be chosen arbitrarily regardless of the functional form of $E(x)$.

To compute correlations, we use \eqref{eq:E_for_arbitrary_x0}, with
$E(x)=\langle E(x)\rangle+\tilde{E}(x)$, to find
\begin{equation}
  C^{E}(x,y)=\langle\tilde{E}(x)\tilde{E}(y)\rangle
  = \langle\tilde{E}_{0}\tilde{E}_{0}\rangle+\langle\tilde{E}_{0}\tilde{E}_{1}(y)\rangle
    +\langle\tilde{E}_{1}(x)\tilde{E}_{0}\rangle+\langle\tilde{E}_{1}(x)\tilde{E}_{1}(y)\rangle.
  \label{eq:C-E-initial-expression}
\end{equation}
Using $\rho_{q}=1-\rho_{e}=-\tilde{\rho}_{e}$, we find 
\begin{equation}
  C_{00}^{E}=\langle\tilde{E}_{0}\tilde{E}_{0}\rangle
  =\int_{0}^{1}zdz\int_{0}^{1}w\langle\tilde{\rho}_{e}(z)\tilde{\rho}_{e}(w)\rangle\,dw;
  \label{eq:C00}
\end{equation}
we also find 
\begin{equation}
  C_{10}^{E}(x)=\langle\tilde{E}_{1}(x)\tilde{E}_{0}\rangle
  =\int_{0}^{1}wdw\int_{0}^{x}\langle\tilde{\rho}_{e}(w)\tilde{\rho}_{e}(z)\rangle dz,
  \label{eq:C10}
\end{equation}
and similarly for $C_{01}(y)$, and (e.g., for $x>y$), 
\begin{equation}
C_{11}^{E}(x,y)=\int_{0}^{x}dz\int_{0}^{y}\langle\tilde{\rho}_{e}(z)\tilde{\rho}_{e}(w)\rangle\,dw.\label{eq:C11}
\end{equation}
Putting these together, we have 
\begin{align}
C^{E}(x,y)= & \int_{0}^{1}zdz\int_{0}^{1}wC(z,w)\,dw+\int_{0}^{1}wdw\int_{0}^{x}C(w,z)dz\nonumber \\
            & +\int_{0}^{1}zdz\int_{0}^{y}C(z,w)dw+\int_{0}^{x}dz\int_{0}^{y}C(z,w)\,dw.
              \label{eq:C-Electric}
\end{align}

For the special $\delta$-function case of Eq.~\eqref{eq:delta-khat}, a substitution into
\eqref{eq:C-Electric} yields 
\begin{equation}
  C^{E}(x,y)=\frac{1}{N_{p}}\left[\text{min}(x,y)-xy
    +\frac{x(x-1)}{2}+\frac{y(y-1)}{2}+\frac{1}{12}\right],
  \label{eq:Full-CE(x,y)}
\end{equation}
which is extended outside $0<x,y<1$ to be periodic in both arguments.
Equation~(\ref{eq:Full-CE(x,y)}) can be cast into the form
\begin{equation}
  C^{E}(x,y)=\frac{1}{2N_{p}}\left[-|x-y|+(x-y)^{2}+\frac{1}{6}\right],
  \label{eq:CE_translation_invariant}
\end{equation}
showing explicitly the translational invariance
$C^{E}(x,y)=C^{E}(x-y)$  and thus
independence of the initial point of integration, $x_{0}$, in addition to
the symmetry $C^E(-x)=C^E(x)$.
The first term in Eq.~(\ref{eq:Full-CE(x,y)}), $\propto\text{min}(x,y)$,
is the Brownian motion result, obtained by assuming a Poisson probability distribution
of particle numbers in each differential region. The sum of the first two terms represents the 
Brownian bridge \cite{Aspects_BB_2008},
a random walk with negative correlations that force the electric field
to be equal at both ends $E(0)=E(1)=0$; the physical origin of this
condition is the net neutrality of the plasma $\int\rho_q(x)dx=0$.
The complete result \eqref{eq:Full-CE(x,y)} using the zero potential assumption
(\ref{eq:ZeroPotential}) can be identified as
the \emph{Ornstein-Uhlenbeck bridge} \cite{Brownian_2nd_order2}.
The three different cases are discussed in more detail in
Appendix~\ref{app:brownian_bridge}.

\begin{figure}[h!]
\centering \includegraphics[width=0.7\textwidth]{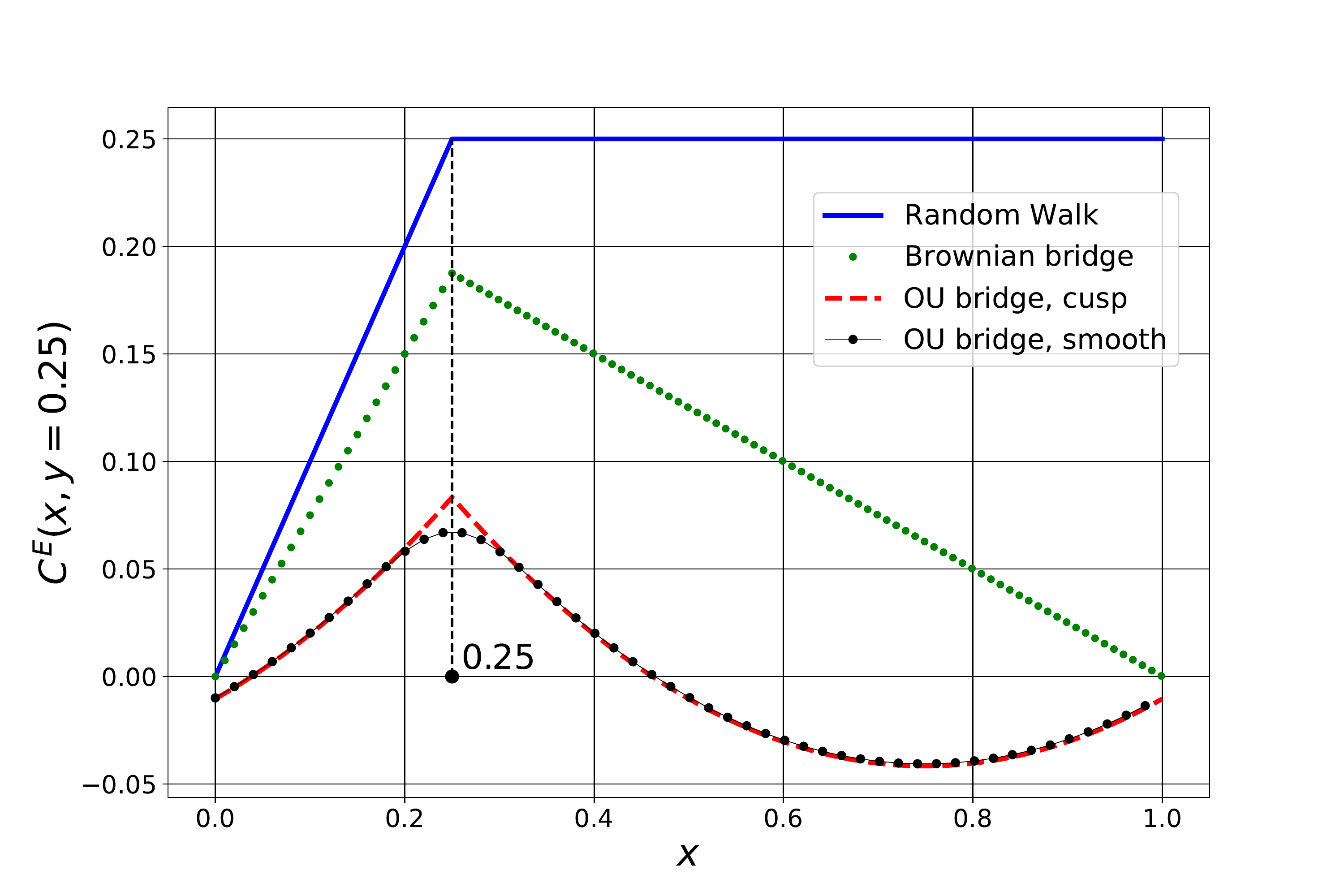}
\caption{Comparison of the electric field covariance matrix $C^E(x,y)$, with $y=0.25$,
for the random walk, the Brownian
bridge, and the Ornstein-Uhlenbeck bridge.
The black curve is the smooth Ornstein-Uhlenbeck bridge with a
linear (tent) kernel
of width $h=0.2$. Note that only the two Ornstein-Uhlenbeck bridge
cases are periodic and have $C^E$  significantly reduced relative to the other two processes.}
\label{fig:BB-plots} 
\end{figure}

The electric field correlations $C^{E}(x-y)$ in Eq.~\eqref{eq:CE_translation_invariant} are 
plotted in Fig.~\ref{fig:BB-plots}
for a fixed value of $y$, showing a maximum at a cusp at $x=y$, with $C^{E}$
being negative over about three parts and positive over about two
parts of the range of $x-y$. Also shown in Fig.~\ref{fig:BB-plots}
is the covariance matrix for the Poisson case (random walk) and for the Brownian
bridge, neither having translational invariance,
both with cusps at $x=y$. Notice that for the Poisson case  $C^{E}(1,y)\neq C^{E}(0,y)$.
For the Brownian bridge case, we have $C^{E}(1,y)=C^{E}(0,y)=0$. It is easy
to show from Eq.~(\ref{eq:Full-CE(x,y)})
that for periodic boundary conditions $C^{E}$ for the Ornstein-Uhlenbeck bridge is maximal
at the cusp at $x=y$ (and at every periodic image), it has a smooth
minimum exactly between the maxima, and has continuous derivatives at the endpoints
$x=0,1$. Notice that the Brownian bridge
correlations are positive, but less than those of the Poisson case
for $x>y$, whereas correlations for the Ornstein-Uhlenbeck case are negative
over an appreciable region of $(x,y)$ and are significantly smaller in magnitude.
Finally, note the aperiodic behavior
of $C^{E}$ for the random walk and the cusps at $x=0$ and $x=1$
for the Brownian bridge. We have also computed $C^{E}(x-y)$ for $\hat{K}$
equal to the linear tent function kernel, which is the convolution of two boxcar kernels of finite
width $h$ (rather than $\hat{K}(x)=\delta(x)$).
The result for $h=0.2$ is shown shown in Fig.~\ref{fig:BB-plots} as
the ``smooth'' Ornstein-Uhlenbeck bridge.

The implications of the lower values of the correlations to particle methods is as follows.
Smaller variances ($x=y$) are desirable because they correspond
to lower noise level. For $x\ne y$, the lower level of the magnitude of the correlations is
expected because of the property $\int\!C^E(x,y)\,dx=0$. Also, in a physical system, i.e., 
in the limit of large $N_p$, such correlations vanish; therefore lower $C^E(x,y)$ is expected to
improve the fidelity of the numerical results.

Recall that $\int\!C(x,y)\,dx=0$ [cf. Eq.~\eqref{eq:intCdy=00003D00003D0}].
A similar general result can be derived for the electric field correlations $C^{E}$.
Indeed, we have 
\begin{equation}
  \int_{0}^{1}C^{E}(x,y)dy=\left<\tilde{E}(x)\int_{0}^{1}\tilde{E}(y)dy\right> = 0,
  \label{eq:int_C^E_zero}
\end{equation}
which vanishes because of the relation \eqref{eq:ZeroPotential}.
In particular, \eqref{eq:int_C^E_zero} can be verified by a direct
calculation for the special case of the covariance matrix \eqref{eq:Full-CE(x,y)}
(or \eqref{eq:CE_translation_invariant}).
As noted for the density covariance matrix, the covariance $C^{E}$ also has
an eigenfunction with eigenvalue
zero, namely $\int_{0}^{1}C^{E}(x,y)u(y)dy=0$ for $u(y)=1$;
we will return to this point in Sec.~\ref{sec:grid_discretization}.

\section{Statistical analysis of error in non-uniform density}
\label{sec:stat_nonuniform}

In this section we discuss a statistical study of non-uniform density
distributions. We now use the more general
term ``error'' or ``statistical error'' instead of ``noise,'' as 
we will show that noise is only part of the total
error, characterized by the \emph{variance},
the other important contribution being the \emph{bias}.

\subsection{Optimal kernel size: bias-variance optimization}
\label{sec:optimal_kernel}

Let us evaluate the mean-square difference (error) $Q$
(henceforth simply error; the actual error can be calculated as $\sqrt{Q}$;)
between the estimated density,
$\rho_{e}(x)$, and the true density $\rho(x)$, where now $\rho(x)$ is not assumed to be constant.
The quantity $Q$ is 
\begin{equation}
  Q=\left<\big(\rho_{e}(x)-\rho(x)\big)^{2}\right>
  =\left<\Big(\rho_{e}(x)-\left<{\rho}_{e}(x)\right>+\left<{\rho}_{e}(x)\right>-\rho(x)\Big)^{2}\right>\,,
  \label{eq:1st-def-Q}
\end{equation}
where we remind the reader that $\langle f\rangle$ is given by
Eq.~\eqref{eq:f_x_ave} or \eqref{eq:f_xy_ave}.
This quantity equals (omitting the argument $x$ for clarity) 
\begin{equation}
  Q = \left<\left(\rho_{e}-\left<{\rho}_{e}\right>\right)^{2}\right>
  +2 \big<\left(\rho_{e} -\left<{\rho}_{e}\right>\right)\left(\left<{\rho}_{e}\right>-\rho\right)\big>
  +\left<\left(\left<{\rho}_{e}\right>-\rho\right)^{2}\right>\,.
  \label{eq:Q-again}
\end{equation}
We recognize that the factor $\left(\left<{\rho}_{e}(x)\right>-\rho(x)\right)$
in the middle and third terms is not a random variable;
since the other factor in the middle term is zero, we find 
\begin{equation}
Q=Q_{1}+Q_{2},\label{eq:Q_total}
\end{equation}
with 
\begin{equation}
  Q_{1}=\left<\rho_{e}^{2}\right>-\left<\rho_{e}\right>^{2},
  \qquad\label{eq:Q1-def}
\end{equation}
\begin{equation}
  Q_{2}= \big(\left<{\rho}_{e}\right>-\rho\big)^{2}.
  \label{eq:Q2-def}
\end{equation}

To proceed, we go back to the Taylor expansion Eq.~\eqref{eq:<rhoe>-2},
  noting that in a non-uniform density $\rho''(x)\ne 0$, and write it as
\begin{equation}
  \langle\rho_{e}(x)\rangle=\rho(x)+B(x)+O(h^{4})
  \label{eq:<rhoe>-with-bias}
\end{equation}
with
\begin{equation}
  B(x)= h^{2}\frac{\rho''(x)}{2}\int\zeta^{2}K_{f}(\zeta)d\zeta\,.
  \label{eq:bias}
\end{equation}
The quantity $B(x)$ is called the statistical \emph{bias}.
Since $\int\! \left<\rho_e(x)\right>\,dx=1$, the bias
satisfies $\int\! B(x)\,dx = 0$, consistent with the periodic
boundary conditions on $\rho(x)$.

For the first term in $Q_{1}$ we have 
\[
\langle\left(\rho_{e}(x)\right)^{2}\rangle=\sum_{\mu\nu}q_{\mu}q_{\nu}\langle K(x-\xi_{\mu})K(x-\xi_{\nu})\rangle
\]
\begin{equation}
  =\sum_{\mu}q_{\mu}^{2}\int K(x-\xi)^{2}\rho(\xi)d\xi
  +\sum_{\mu\ne\nu}q_{\mu}q_{\nu}\left(\int K(x-\xi)\rho(\xi)d\xi\right)^{2}\,,
  \label{eq:1st-part-of-Q1}
\end{equation}
where we have again split the sums into diagonal terms ($\mu=\nu$) and
off-diagonal terms ($\mu\neq\nu$). Again, changing variables
and Taylor expanding, we find 
\begin{align}
  \langle\left(\rho_{e}(x)\right)^{2}\rangle
   = &\,\, \frac{\rho(x)}{N_{p}h}\int K_{f}(\zeta)^{2}d\zeta+O\left(\frac{h}{N_{p}}\right)\nonumber \\
    & +\left(1-\frac{1}{N_{p}}\right)\langle\rho_{e}(x)\rangle^{2} \nonumber\\
   \approx &\,\, \frac{\rho(x)}{N_{p}h}\int K_{f}(\zeta)^{2}d\zeta
           +\langle\rho_{e}(x)\rangle^{2} - \frac{1}{N_p}\left(\rho(x)+B(x) \right)^2  \nonumber\\
  = &\,\, \frac{\rho(x)}{N_{p}h}\int K_{f}(\zeta)^{2}d\zeta
      +\langle\rho_{e}(x)\rangle^{2} - \frac{1}{N_p}\rho(x)^2\,,
      \label{eq:1st-part-of-Q1=00003D00003Dagain}
\end{align}
using Eq.~\eqref{eq:<rhoe>-with-bias} 
and neglecting terms of first and higher orders in $h/N_p$.
Note that the $-\rho(x)^2/N_p$ term in \eqref{eq:1st-part-of-Q1=00003D00003Dagain}
does \emph{not} arise from the Taylor expansion \eqref{eq:<rhoe>-2}.
For the purpose of the present argument we neglect that term
since $\rho(x)^2$ is of order one and $N_p$ is typically a large number
in particle simulations.
However, recall that this is the same factor responsible for the negative
correlations in Sec.~\ref{sec:stat_uniform}, where although small,
it had a non-negligible cumulative effect; we will
revisit its importance in Sec.~\ref{sec:numerical}.

We find that the $\left<\rho_{e}(x)\right>^{2}$ 
terms cancel in Eqs.~\eqref{eq:Q1-def} and we are left with 
\begin{equation}
  Q_{1}=\frac{\rho(x)}{N_{p}h}\int K_{f}(\zeta)^{2}d\zeta.
  \label{eq:Q1=00003D00003Dfinal}
\end{equation}
We also have from Eq.~\eqref{eq:Q2-def}, \eqref{eq:<rhoe>-with-bias}, and \eqref{eq:bias}
\begin{equation}
  Q_{2}=B(x)^2=\left(h^{2}\frac{\rho''(x)}{2}\int\zeta^{2}K_{f}(\zeta)d\zeta\right)^{2}.
  \label{eq:Q2=00003D00003Dfinal}
\end{equation}
Eqs.~\eqref{eq:Q1=00003D00003Dfinal}, \eqref{eq:Q2=00003D00003Dfinal} give
\begin{equation}
  Q=Q_{1}+Q_{2}=\frac{\rho(x)}{N_{p}h}\int K_{f}(\zeta)^{2}d\zeta
  +\left(h^{2}\frac{\rho''(x)}{2}\int\zeta^{2}K_{f}(\zeta)d\zeta\right)^{2}.
  \label{eq:V+B^2}
\end{equation}
The first term, $Q_{1}=V$, is the \emph{variance} (the diagonal terms
in the covariance matrix) and the second term, $Q_{2}$, is the square
of the \emph{bias}, $Q_{2}=B^{2}$. (Note that the addition of the
bias to $\rho(x)$ in Eq.~(\ref{eq:<rhoe>-with-bias}) is analogous
to the smoothing obtained by diffusion of the density over a time
interval $t$, $\rho(x)\rightarrow\left(1+Dt\partial_{x}^{2}+\cdots\right)\rho(x)$,
where $D$ is a diffusion coefficient and $h^{2}\int\zeta^{2}K_{f}(\zeta)d\zeta/2\rightarrow Dt$.)
Writing 
\begin{equation}
  C_{1}=\int_{-1/2}^{1/2} K_{f}(\zeta)^{2}d\zeta\,,\qquad C_{2}=\int_{-1/2}^{1/2}\zeta^{2}K_{f}(\zeta)d\zeta\,,
  \label{eq:C_1_C_2}
\end{equation}
we have
\begin{equation}
  Q=V+B^{2}=\frac{\rho(x)C_{1}}{N_{p}}\frac{1}{h}+\frac{\rho''(x)^{2}C_{2}^{2}}{4}h^{4}.
  \label{eq:Q=00003D00003DV+B^2}
\end{equation}
Clearly the factors $C_1$ and $C_2$ are related to the kernel shape, whereas the kernel width is
represented by $h$.
The interpretation of the two contributions in the result Eq.~\eqref{eq:Q=00003D00003DV+B^2} is
as follows: the bias is an error caused by estimating the spatially varying density
$\rho(x)$ using a kernel of width $h$, i.e., it is a \emph{finite size} particle effect;
the variance is an error (noise) due to the \emph{finite number} of particles.
The balance between the two effects is reached when
\begin{equation}
  \frac{1}{\rho(x) N_p h} \sim \frac{\rho''(x)^2}{\rho(x)^2}h^4 \sim \left(\frac{h}{l}\right)^4\,,
  \label{eq:BV_balance}
\end{equation}
where we have defined the density gradient length scale $l=\sqrt{\rho(x)/|\rho''(x)|}$ and have assumed
the shape coefficients $C_1$ and $C_2$ are of order unity. We see that the bias error term dominates
for $h$ large compared to $l$ (more smoothing of the density); more specifically, when 
\begin{equation}
  \left(\frac{l}{h}\right)^4 \ll {\rho(x) N_p h} \equiv N_h\,.
  \label{eq:BV_balance_Nh}
\end{equation}
One recognizes the product $N_h= \rho(x)N_p h$ as the
typical number of particles within the kernel width $h$.
The variance error dominates when the opposite inequality holds.
The condition \eqref{eq:BV_balance} will be revisited
in Sec.~\ref{sec:numerical} where numerical examples are presented.

For a more quantitative description, optimizing over $h$ for fixed $x$, we find a \emph{minimum} at 
\begin{align}
  h  =\Hopt & =\left(\frac{\rho(x)C_{1}}{N_{p}\rho''(x)^{2}C_{2}^{2}}\right)^{1/5}, \label{eq:h-opt}\\
  \Qmin &= \frac{5}{4}\left(\frac{\rho(x)|\rho''(x)|^{1/2}C_{1}C_{2}^{1/2}}{N_{p}}\right)^{4/5},
          \label{eq:Qmin} \\
  Q''(\Hopt) &= 5 \left( \frac{\rho(x) |\rho''(x)|^3 C_1C_2^3}{N_p} \right)^{2/5}\,. \label{eq:d2Qmin}
\end{align}
This process of minimizing $Q$ is called \emph{bias-variance optimization}
\cite{bishop_pattern_2006,Sammut_encyclopedia_2017}.
We see that the very factor that has made particle methods so useful---the finite size of
computational particles---is not without its drawbacks, leading to the
bias error in the density estimation. However, our result provides a guideline
for taking advantage of this factor as it varies \emph{oppositely} to the other error contribution,
that of the variance (noise). Thus we arrive at the trade-off between bias and variance error
embodied in the BVO process just described.

Eqs.~\eqref{eq:h-opt} and \eqref{eq:Qmin} suggest that the optimal value of $h$ depends on
$x$. A reasonable alternative is to let $\rho(x)\rightarrow\int_{0}^{1}\rho(x)dx=1$
and $\rho''(x)^{2}\rightarrow\int_{0}^{1}\rho''(x)^{2}dx$, i.e., to integrate Eq.~(\ref{eq:Q=00003D00003DV+B^2}),
leading to the \emph{mean integrated square error} result 
\begin{align}
  h=\HoptA &= \left(\frac{C_{1}}{N_{p}\left(\int\rho''(x)^{2}dx\right)C_{2}^{2}}\right)^{1/5},
  \label{eq:h-opt-1} \\
  \QminA & =\frac{5}{4}\left(\frac{\left(\int\!dx\,|\rho''(x)|^{2}\right)^{1/4}C_{1}C_{2}^{1/2}}{N_{p}}\right)^{4/5},
  \label{eq:MISE-Qmin} \\
  Q''(\HoptA) &= 5 \left( \frac{(\int(\rho''(x))^2 dx)^{3/2} C_1C_2^3}{N_p} \right)^{2/5}\,. \label{eq:MISE-d2Qmin}
\end{align}
If $|\rho''(x)|$ does not 
vary too much, it is possible to take advantage
of the fractional power in Eq.~(\ref{eq:MISE-Qmin}) to use a kernel
of width $h\approx \HoptA$
  throughout the whole simulation domain. This is especially important in
cases in which a choice is made to have a fixed relation between the
kernel width $h$ and a uniform grid spacing $\Delta$.

Note the dependence of the quantities in Eqs.~\eqref{eq:h-opt}, \eqref{eq:Qmin}, and
\eqref{eq:d2Qmin} on $N_p$, namely $\Hopt\propto N_p^{-1/5}$, $\Qmin\propto N_p^{-4/5}$ and
$Q''(\Hopt)\propto N_p^{-2/5}$. This shows that $\Hopt$ is quite insensitive
to the number of particles. It is interesting to note that
$\Qmin$ has a slightly weaker scaling that the usual $\propto N_p^{-1}$
scaling of the variance alone, implying scaling $1/N_p^{2/5}$ vs. $1/\sqrt{N_p}$ for the
error $\propto\Qmin^{1/2}$.

\begin{table}
  \footnotesize
\centering 
\begin{tabular}{l|c|c|c|c}
  \hline \rule{0pt}{3ex}
  Kernel  & $C_{1}$  & $C_{2}$ & $\left(C_1 C_2^{1/2}\,\right)^{4/5}$ & $\left(C_1/C_2^2\right)^{1/5}$ \\[0.5ex]
  \hline \hline \rule{0pt}{2ex}
  Boxcar  & $1$  & $1/12$  & $0.370$ & $2.70$ \\
  \hline \rule{0pt}{2ex}
  Linear (tent) & $4/3$ & $1/24$ & $0.353$ & $3.78$ \\
  \hline \rule{0pt}{2ex}
  Quadratic & $33/20$ & $1/36$ & $0.356$ & $4.63$ \\
  \hline \rule{0pt}{2ex}
  Trapezoidal & $5/4$ & $5/108$ & $0.350$ & $3.57$ \\
  \hline \rule{0pt}{2ex}
  Epanechnikov  & $6/5$  & $1/20$ & $0.349$ & $3.44$ \\
  \hline\hline
\end{tabular}
\caption{The values of the coefficients $C_1$ and $C_2$ for the fundamental kernels in
  Table~\ref{table:fundamental_kernels} and Fig.~\ref{fig:fundamental_kernels},
  including those used in Fig.~\ref{fig:sketch_BVO}.
  The quantity in column $3$ is the factor appearing in $\Qmin$.
  The quantity in column $4$ is the factor appearing in $\Hopt$ as well as in the width
  $W_Q$ from Eq.~\eqref{eq:W_Q}. Note that  column $3$ varies little between the kernels,
  but column $4$ varies by almost a factor of two.
}
\label{table:Kernel_C_1_C_2}
\end{table}

Kernels with compact support are typically used in particle simulations,
for computational efficiency. Among all kernels with compact support, with
width equal to one, and 
having $\int K_{f}(\zeta)d\zeta=1$, the Epanechnikov kernel
minimizes the factor $C_{1}C_{2}^{1/2}$
in $\Qmin$ \cite{epanechnikov_non-parametric_1969,Li_nonparametric_2006}.
However, the factor $(C_{1}C_{2}^{1/2})^{4/5}$ in $\Qmin$
varies little between different kernels, 
so that the kernel \emph{shape} has little influence on $\Qmin$.

\begin{figure}[!t]
  \begin{center}
\includegraphics[width=0.7\textwidth]{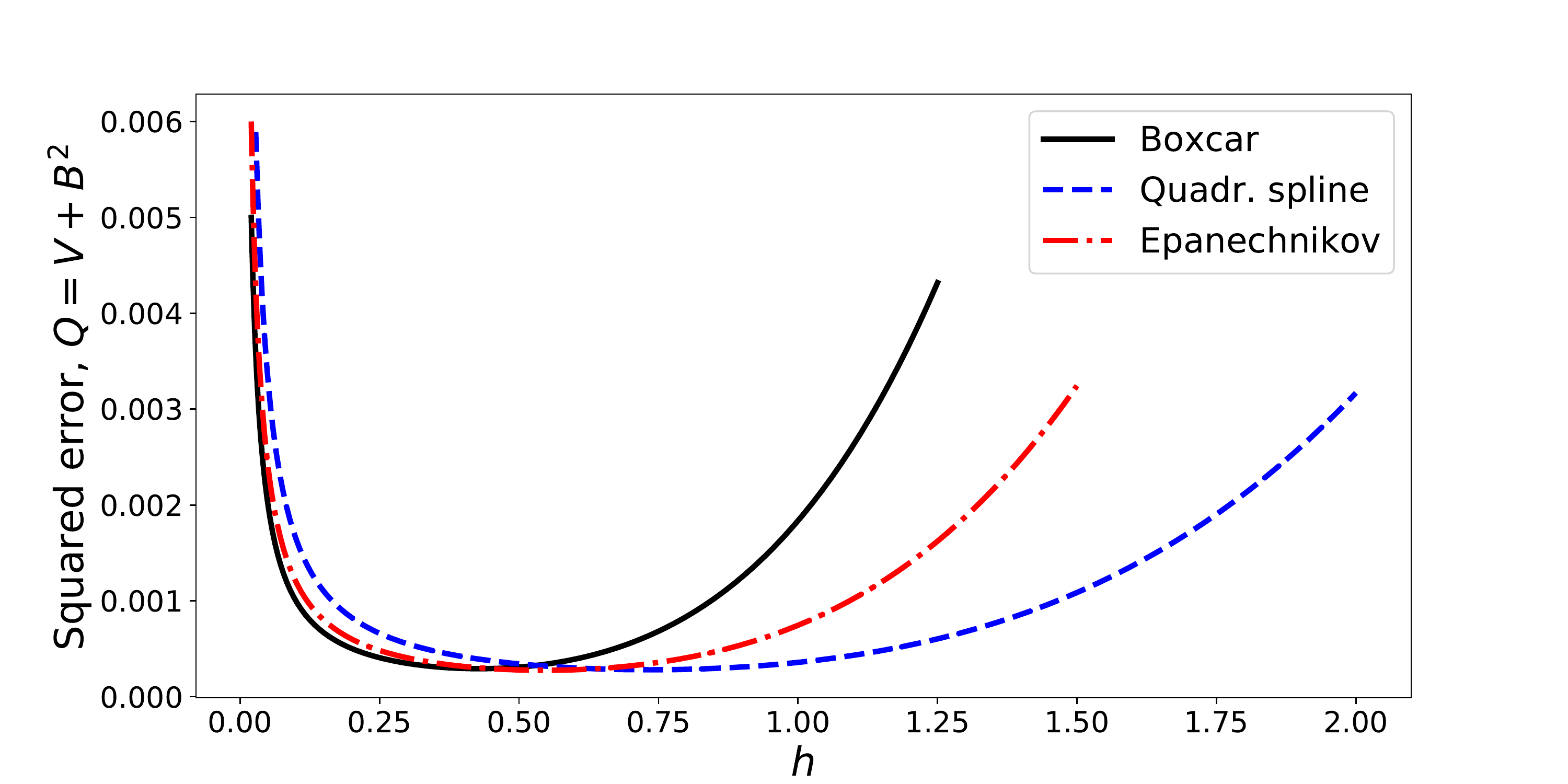}
  \end{center}
  \caption{A sketch of the error, Eqs.~\eqref{eq:Q=00003D00003DV+B^2}, with $\rho(x)$ and 
  $\rho''(x)^2$ averaged, as a function
  of $h$ for the boxcar, Epanechnikov and quadratic spline kernels,
    showing a slightly lower minimum at $\Hopt$ for the Epanechnikov. 
    Also, note the higher values of $\Hopt$ and the 
    broader minima for the Epanechnikov and quadratic spline kernels,
    in agreement with the results summarized in Table~\ref{table:Kernel_C_1_C_2}.}
  \label{fig:sketch_BVO}
\end{figure}

A plot of the $Q$ vs. $h$ [cf.~Eq.~\eqref{eq:MISE-Qmin}]
is shown in Fig.~\ref{fig:sketch_BVO}, using the mean integrated square error
approximation and $\int\rho(x)dx = \int\rho''(x)^2 dx = 1$.
The three curves $Q(h)$ correspond to
the boxcar, quadratic, and Epanechnikov kernels, 
defined in Table~\ref{table:fundamental_kernels}.
The coefficients $C_1$ and $C_2$, calculated from
Eq.~\eqref{eq:C_1_C_2}, are given in the first two columns of Table~\ref{table:Kernel_C_1_C_2}.
The number of particles is taken to be $N_p=10^4$.
The range of $h$ is chosen so that the sections dominated by variance
(small $h$) and by bias (large $h$) are clearly seen,
as well as the intermediate values where a minimum is attained.
Note the shape dependencies $\HoptA\propto (C_1/C_2^2)^{1/5}$,
$\Qmin\propto (C_1 C_2^{1/2})^{4/5}$.
The width of the minimum of $Q(h)$ is proportional to
\begin{equation}
  W_{\rm Q}\propto (\Qmin/Q''(\Hopt))^{1/2} \propto (C_1 / C_2^2)^{1/5},
  \label{eq:W_Q}
\end{equation}
which is the same factor appearing in the expressions for $\Hopt$.
These combinations are also listed in Table~\ref{table:Kernel_C_1_C_2}.
The values in the third column, i.e., $(C_1C_2^{1/2})^{4/5}$,
confirm that the shape has a minimal effect on $\Qmin$
and is consistent with the slightly lower minimum of the Epanechnikov kernel
compared to the other two kernels in Fig.~\ref{fig:sketch_BVO},
as discussed above.
It is also easy to see that the location of $\Hopt$ for the three curves
in Fig.~\ref{fig:sketch_BVO}
is consistent with the values in the last column of that table;
for example, note that the ratio of the values of $\HoptA$ for the Epanechnikov kernel
to the boxcar kernel, equal to $3.44/2.70\approx 1.27$.
Also, note that the last column in Table~\ref{table:Kernel_C_1_C_2}
is the factor entering in the width $W_{\rm Q}$.

An important feature of the error curve is the relatively broad minimum,
which means $\Qmin$ does not vary significantly over a relatively large
range of $h$ values around its optimal value. This has the practical
consequence that when $\rho(x)$ and $\rho''(x)$ have a modest variation over
the simulation domain, a fixed kernel of width close to $\HoptA$
still provides a near optimal density estimate; hence, we have another justification
for using the averaged quantities, Eqs.~\eqref{eq:h-opt-1} and \eqref{eq:MISE-Qmin}.
Fig.~\ref{fig:sketch_BVO} shows that the minimum for the 
Epanechnikov kernel is broader by less than a factor $2$ than that for the boxcar kernel,
while for the quadratic spline that factor is closer to $2$;
both of these observations are consistent with the values in the
last column of Table~\ref{table:Kernel_C_1_C_2}.
Quantitative comparison of Eqs.~\eqref{eq:h-opt}--\eqref{eq:MISE-Qmin}
against numerical simulations is done in Sec.~\ref{sec:numerical},
which relaxes the approximations of this analysis.

Next we illustrate how our results can be applied to
algorithms of practical importance, i.e., including grid discretization.

\section{Grid discretization}
\label{sec:grid_discretization}

So far we have obtained results in continuous spatial variables.
For numerical purposes, we need to perform grid discretization.
Clearly, there is no
universal discretization and different problems may benefit
from different discretizations. For illustration purposes,
in this section we choose a particular finite difference
discretization that is commonly used in electrostatic PIC algorithms
\cite{birdsall_plasma_2004} but repeating the analysis
for other discretizations is straightforward,
including the use of finite elements.

\subsection{Estimation kernel, particle shapes, and the sum rule}
\label{sec:sum_rule}

An important step in obtaining a complete particle algorithm is the
connection of Lagrangian particles with a Eulerian grid.
This connection is given by a \emph{charge deposition rule} and
traditionally done with so-called spline
functions \cite{birdsall_plasma_2004}.
Although spline functions of varying degree of smoothness and width are available,
they have the following two limitations:
(i) their width is an integer number
of cell widths, $i\Delta$, $i=1,2,\ldots$; and (ii) their width and smoothness
are strictly related, with smoother particles being wider.
The smoothness of a particle becomes important
  when force interpolation from the grid to the particle position, especially
  when a particle crosses cell boundaries. Since in this work we are
  not addressing the full PIC cycle, we emphasize the importance of
  particle width over smoothness. As well,
if one were to use particles arbitrarily related to
the grid spacing $\Delta$ to minimize noise and error,
per our theoretical developments, the above restrictions may present
a drawback. For example, when high grid resolution is desired (small $\Delta$)
while $\Hopt$ is (relatively) large, that would require a particle
that spans a large number of cells. If splines are used, they would
be of high order and computationally expensive because of the larger
number of floating point operations associated with
high order polynomials. In fact, this is why practitioners rarely go beyond
fourth order spline functions. The advantage of being able to choose separately
particle smoothness and width becomes obvious.

In this section we address the relaxation of the two limitations
discussed above, those associated with the smoothness and the width of particle shapes.
The former was discussed in Ref.~\cite{evstatiev_variational_2013},
where the \emph{smoothness} of particle shapes was decoupled from its width.
An example of a cubic particle shape depositing charge on three
grid points (same as the quadratic spline) was given therein;
by the same method, for example, one could devise a quadratic particle wider than
three cells, etc. The key element
that allows this generalization is the \emph{distinction} between the
kernel $K(x)$ and the particle shape (factor) $S(x)$, 
alluded to in Sec.~\ref{sec:density_estimation}. 
Therefore, before proceeding to address the limitation associated with
the particle width, we return to a discussion of the difference between $K(x)$ and $S(x)$.

It is the sum rule property that distinguishes a kernel from a particle shape, see 
Eq.~\eqref{eq:S_sum_rule}. We will require that a particle shape satisfies the sum rule,
whereas we will not impose this requirement on a kernel.
That property states that the sum of the fractions of the computational
particle's charge deposited on the grid sum exactly to the charge
carried by the particle, and this is true at any (continuously varying)
particle position $\xi$. As a consequence, the total charge
of the system \emph{after being deposited on the
grid} is also conserved. For example, for density that integrates to unity
on a uniform grid with spacing $\Delta$, we have 
$\int_{0}^{1}\!\rho_{e}(x)\,dx=\sum_{\mu=1}^{N_{p}}q_{\mu}=1\,.$
Consider the amount of charge a single particle deposits on the grid point $x_i$.
We use a charge deposition rule based on a particle shape $S(x)$, which gives for the
fraction of that charge $ q_\mu \Delta S(x_i-\xi_\mu)$.
We find that the density on the grid point $x_i$ due to depositing the charge from all the particles is
\begin{equation}
\rho_{e}(x_i)\equiv \rho_{e,i} = \sum_{\mu=1}^{N_p}q_\mu S(x_i-\xi_\mu)\,.
  \label{eq:rho_k}
\end{equation}
For particles with equal charges, $q_{\mu}=1/N_{p}$, we obtain
\begin{equation}
  \sum_{i=1}^{N_{g}}\Delta\,\rho_{e,i}
  =\sum_{i=1}^{N_{g}}\Delta \sum_{\mu=1}^{N_{p}}q_{\mu}S_{i}(\xi_{\mu})
  =\sum_{\mu=1}^{N_{p}}\frac{1}{N_{p}}\sum_{i=1}^{N_{g}}\Delta S_{i}(\xi_{\mu})
  =\sum_{\mu=1}^{N_{p}}\frac{1}{N_{p}}=1\,.
\end{equation}
We have defined $S_k(\xi_\mu)\equiv S(x_k-\xi_\mu)$ and 
invoked the sum rule, Eq.~\eqref{eq:S_sum_rule}. Thus, the sum rule implies that
the total charge assigned to the grid is preserved.
If a particle shape does not satisfy the sum rule (i.e. it obeys conditions
\eqref{eq:S_normalization-1}--\eqref{eq:S_finite-1} for a kernel but not the sum rule),
the lack of exact charge conservation would allow $E(1) \neq E(0)$.
Of course, the total charge of the
system is always conserved as long as no association with a computational
grid is made, as discussed in Sec.~\ref{sec:density_estimation}.
Examples of particle shapes (satisfying the sum rule) are given in
Table~\ref{table:particle_shapes}. The first three are familiar from
Ref.~\cite{birdsall_plasma_2004}; the last (trapezoidal) particle shape is discussed below.
The charge deposition on the grid point $x_i$, associated with each of the
particles in Table~\ref{table:particle_shapes}, is found by the substitution $x=\xi_\mu-x_i$,
for $i=1,2,\ldots, N_g$.

\begin{table}
  \footnotesize
  \begin{center}
    \begin{tabular}{l|l}
      \hline \rule{0pt}{3ex} 
    Particle shape & Definition   \\
    \hline\hline \rule{0pt}{5ex} 
    Boxcar (NGP) & $ S_{B}(x) =  \frac{1}{\Delta}\left\{
                       \begin{tabular}{ll}
                         \ensuremath{1},    \ensuremath{\left|\frac{x}{\Delta}\right|\le\frac{1}{2}}\\[1.25ex]
                         \ensuremath{0}   otherwise\,. 
                       \end{tabular}\right. $  \\[1.25em]
    \hline \rule{0pt}{5ex} 
    Linear spline & $ S_{L}(x) =  \frac{1}{\Delta}\left\{
                    \begin{tabular}{ll}
                      \ensuremath{1-\left|\frac{x}{\Delta}\right|}, \ensuremath{\left|\frac{x}{\Delta}\right|\le 1}\\[1.25ex]
                      \ensuremath{0}    otherwise\,. 
                    \end{tabular}\right. $ \\[1.25em]
    \hline \rule{0pt}{7ex}
    Quadratic spline &  $ S_{Q}(x) =\frac{1}{\Delta}\left\{
                \begin{tabular}{ll}
                  \ensuremath{\frac{3}{4}-\left(\frac{x}{\Delta}\right)^{2}}, \ensuremath{\left|\frac{x}{\Delta}\right|\le1/2}\\[1.25ex]
                  \ensuremath{\frac{1}{2}\left(\frac{3}{2}-\left|\frac{x}{\Delta}\right|\right)^{2}},  \ensuremath{1/2\le\left|\frac{x}{\Delta}\right|\le3/2} \\[1.25ex]
                  \ensuremath{0}    otherwise\,.
                \end{tabular}\right. $  \\[2em]
\hline \rule{0pt}{7ex} 
    Trapezoidal &  $ S_{T}(x) = \frac{1}{\Delta}\left\{
                  \begin{tabular}{ll}
                    \ensuremath{\frac{1}{2}},  \ensuremath{\left|\frac{x}{\Delta}\right|\le1/2}\\[1.25ex]
                    \ensuremath{\frac{1}{2}\left(\frac{3}{2}-\left|\frac{x}{\Delta}\right|\right)},  \ensuremath{1/2\le\left|\frac{x}{\Delta}\right|\le3/2} \\[1.25ex]
                    \ensuremath{0}    otherwise\,, 
                  \end{tabular}\right. $  \\[2em]
    \hline\hline
  \end{tabular}
\end{center}
\caption{Examples of particle shapes. These shapes are similar to the kernels of 
    Table~\ref{table:fundamental_kernels} and Fig.~\ref{fig:fundamental_kernels}, 
    but have integer valued cell width and satisfy the sum rule, Eq.~\eqref{eq:S_sum_rule}.}
\label{table:particle_shapes}
\end{table}

We now describe the generalization associated with particle \emph{width}:
a particle shape satisfying the sum rule \eqref{eq:S_sum_rule} is \emph{not} 
required to have a width equal to an integer number of grid cells.
In fact, such a shape can have (almost -- see below) completely arbitrary width relative to
the grid. To see this, consider a known ``primary'' particle shape that
satisfies properties \eqref{eq:S_normalization-1}--\eqref{eq:S_finite-1}
and the sum rule \eqref{eq:S_sum_rule}, say $S_{0}(x)$, with $\sum_{i}\Delta S_{0}(x_{i}-\xi)=1$
for any value of $\xi\in [0,1]$. Then for an arbitrary kernel $K(x)$ satisfying properties
\eqref{eq:S_normalization-1}--\eqref{eq:S_finite-1}, we perform the \emph{convolution} 
\begin{equation}
  S(x)=\int\!K(y)S_{0}(x-y)\,dy.
  \label{eq:K_convolution}
\end{equation}
The so-obtained new particle shape $S(x)$ satisfies the normalization
\eqref{eq:S_normalization-1}:
\begin{equation}
\int_{0}^{1}\!S(x)\,dx=\int\!K(y)\,dy\int\!S_{0}(x-y)\,dx=\int\!K(y)\,dy=1\,.
\end{equation}
The sum rule \eqref{eq:S_sum_rule} is also easily verified: 
\begin{equation}
  \sum_{i=1}^{N_{g}}\Delta S(x_{i})=\int\!K(y)\sum_{i}\Delta S_{0}(x_{i}-y)\,dy
  =\int\!K(y)\,dy=1\,.
\end{equation}
It is easy to verify that properties Eq.~\eqref{eq:S_symmetry-1}--\eqref{eq:S_finite-1}
are inherited by $S(x)$ as well. The \emph{width} of $S(x)$ equals
to the sum of the widths of $S_{0}(x)$ and $K(x)$. The procedure
just described allows a kernel $K(x)$ of arbitrary width; we conclude
that this construction allows one to generate arbitrary width particle shapes satisfying
the sum rule, including such that are non-integer number of cells
wide. The only condition on the width of $S(x)$ is that it cannot
be less than the width of $S_{0}(x)$, hence the qualifier ``almost'' above;
this is usually not a limitation. We stress that if
a particle shape $S(x)$ is obtained from a kernel $K(x)$ and another
particle shape, $S_{0}(x)$, their functional forms are different
(in addition to their widths being different).

We note that the convolution described by Eq.~\eqref{eq:K_convolution}
is the easiest way to obtain a particle shape that satisfies the sum rule.
However, this is a sufficient but not necessary condition.
As well, choosing a familiar particle shape obeying the sum rule as a primary, $S_0$, is
the easiest way to ensure $S(x)$ satisfies the sum rule; again, this choice
is sufficient but not necessary. In other words, it is possible that other
methods of obtaining particle shapes that satisfy the sum rule exist.

The examples in Table~\ref{table:particle_shapes} satisfy the sum rule
and can either be used as primary to obtain other particle shapes or
directly in a simulation. In fact, all particles from
Table~\ref{table:particle_shapes} can in turn be obtained by the 
convolution formula \eqref{eq:K_convolution} as follows: the boxcar shape is
the convolution of itself with a delta-function kernel;
the linear shape is the convolution of the boxcar shape with a
boxcar kernel of width $\Delta$;
the quadratic spline shape is the convolution of the boxcar shape with a linear kernel
(tent function) of width $2\Delta$; and finally, the trapezoidal shape is
the convolution of the boxcar shape and a boxcar kernel of width $2\Delta$.

\begin{figure}[t!]
  \begin{center}
    \includegraphics[width=0.7\textwidth]{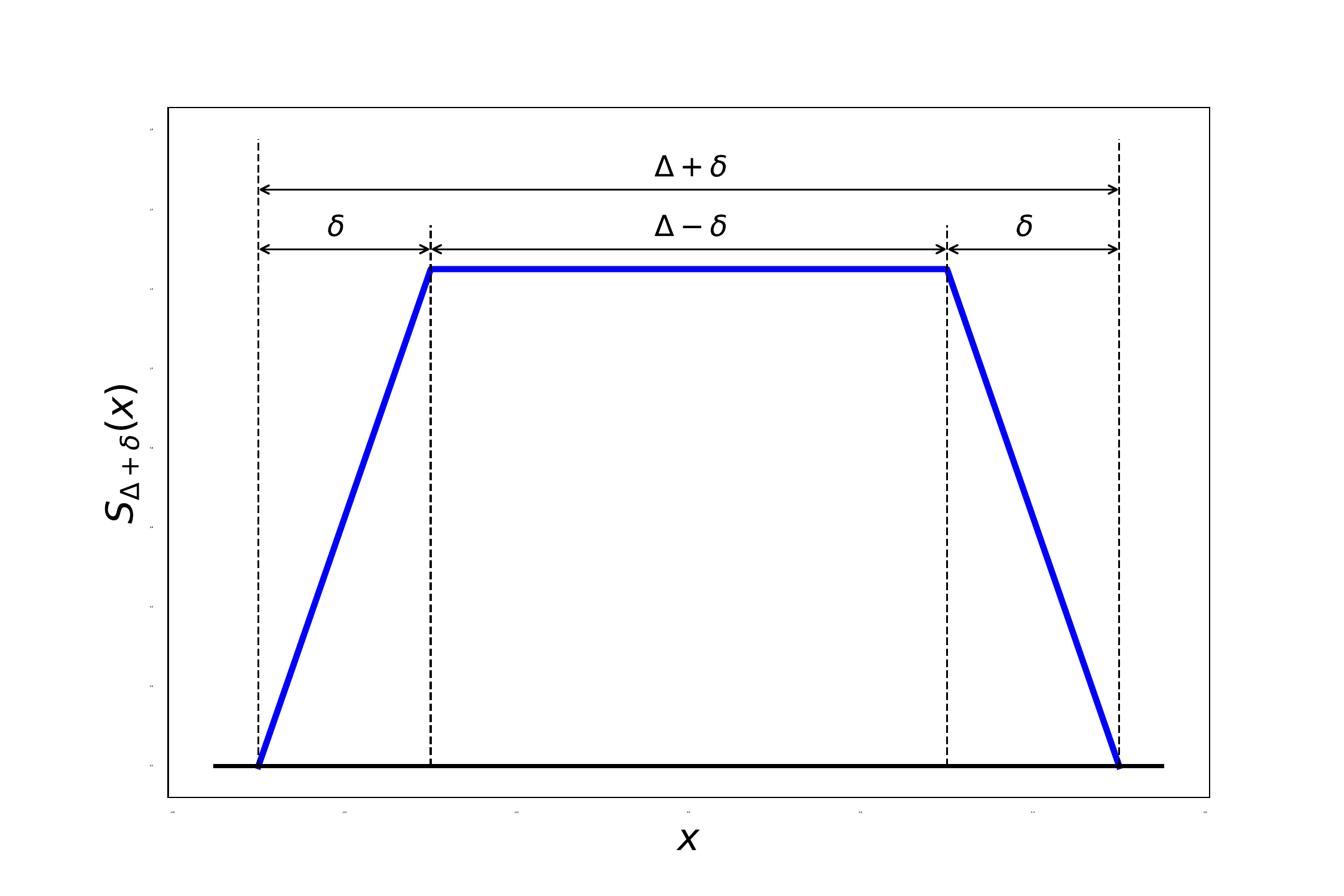}
  \end{center}
  \caption{Illustration of the fractional width particle shape, Eq.~\eqref{eq:S_Delta_h},
    with $\delta/\Delta=0.4$ and $h = 1.4\Delta$.
    Because this quantity is a convolution of two functions, one of which satisfies the sum rule in
      Eq.~\eqref{eq:S_sum_rule}, it satisfies the sum rule for an arbitrary value of $\delta/\Delta$,
      and therefore for an arbitrary value of $h/\Delta<1$.}
  \label{fig:fractional_shape}
\end{figure}

An example of a particle shape that is non-integer number of cells wide is given next.
Consider the convolution of the boxcar particle shape of width $\Delta$ from
Table~\ref{table:particle_shapes} with the boxcar \emph{kernel} of width $\delta$
with $0\le \delta\le \Delta$,
\begin{align}
 K_{\delta}(x) = & \frac{1}{\delta}
 \left\{
 \begin{tabular}{ll}
 \ensuremath{1},  &  \ensuremath{\left|\frac{x}{\delta}\right|\le\frac{1}{2}}\\[1.25ex]
 \ensuremath{0}  &  otherwise\,. 
 \end{tabular}
 \right.
\label{eq:K_boxcar_h}
\end{align}
The convolution formula \eqref{eq:K_convolution} gives the following
\emph{trapezoidal} particle shape, of width $\Delta+\delta$:
\begin{align}
  S_{\Delta+\delta}(x)
  = & \int\!\!  K_\delta(y) S_B(x-y)\,dy \nonumber \\
  = & \frac{1}{\Delta}
      \left\{
      \begin{tabular}{ll}
        $1$, & $0 \le |x| \le \frac{\Delta-\delta}{2}$ \\ [1.25ex]
        $\frac{1}{\delta}\frac{\Delta+\delta}{2}-|\frac{x}{\delta}|$, & $\frac{\Delta-\delta}{2} \le |x| \le \frac{\Delta+\delta}{2}$ \\ [1.25ex]
        $0$, & otherwise\,.
      \end{tabular}
               \right.
  \label{eq:S_Delta_h}
\end{align}
The particle shape \eqref{eq:S_Delta_h} transforms into
the usual $\Delta$-wide boxcar shape in the limit $\delta\rightarrow 0$
and into the usual $2\Delta$-wide linear shape in the limit $\delta \rightarrow \Delta$.
Note that $\int\!\! S_{\Delta+\delta}(x)\,dx = 1$, as expected.
The fractional width particle is illustrated in Fig.~\ref{fig:fractional_shape}.
The charge deposition rule is found by substitution of $\xi-x_i$ into
\eqref{eq:S_Delta_h}, where $x_i$ is the nearest grid point to the particle position
and $|\xi-x_i|\le \Delta/2$. The result is given in Table~\ref{table:fractional_charge_deposition}
and the construction by means of a convolution assures that the sum rule is satisfied 
for arbitrary values of $\delta$.

We remark that the dependence of the particle shape \eqref{eq:S_Delta_h} on the
fractional width $\delta$ is not a scaling transformation via a fundamental
kernel such as changing $h$ in \eqref{eq:K_definition}. Instead, this is a
parametric \emph{shape transformation} (with parameter $\delta$).
Nevertheless, changing the shape via $\delta$ also changes the support of $S_{\Delta+\delta}$
and provides another means of attaining the optimal width, $\Hopt = \Delta+\delta$.

The existence of the fractional width particle shape \eqref{eq:S_Delta_h} was noted
in Ref.~\cite{birdsall_clouds_1969} and has been previously known in particle hydrodynamics.
Its derivation, however, has been based on area weighting arguments and not on the more general
convolution method given by \eqref{eq:K_convolution}. Again, the fractional width trapezoidal
shape \eqref{eq:S_Delta_h} is \emph{not} the same as the trapezoidal shape listed
in Table~\ref{table:particle_shapes}.

 \begin{table}[!t]
   \footnotesize
   \begin{center}
     \begin{tabular}{l|l}
       \hline \rule{0pt}{3ex} 
       Charge deposition rule & Range \\
       \hline\hline \rule{0pt}{3ex} 
       $S_{i-1}(\xi) =  0$ &\\ [1.25ex]
       $S_i(\xi) = \frac{1}{\Delta}$ & $0\le |\xi-x_i|\le \frac{\Delta-\delta}{2}$ \\ [1.25ex]
       $S_{i+1}(\xi) =  0$ & \\ [1.25ex]
       \hline \rule{0pt}{4ex} 
       $S_{i-1}(\xi) =  \frac{1}{\Delta \delta}\left[-\frac{\Delta-\delta}{2}-(\xi-x_i)\right] $ &\\ [1.25ex]
       $S_i(\xi) = \frac{1}{\Delta \delta}\left[\frac{\Delta+\delta}{2}+(\xi-x_i)\right]$ & $-\frac{\Delta}{2}\le \xi-x_i\le -\frac{\Delta-\delta}{2}$ \\ [1.25ex]
       $S_{i+1}(\xi) =  0$ & \\ [1.25ex]
       \hline \rule{0pt}{3ex} 
       $S_{i-1}(\xi) = 0  $ &\\
       $S_i(\xi) = \frac{1}{\Delta \delta}\left[\frac{\Delta+\delta}{2}-(\xi-x_i)\right]$ & $\frac{\Delta-\delta}{2}\le \xi-x_i\le \frac{\Delta}{2}$ \\ [1.25ex]
       $S_{i+1}(\xi) =  \frac{1}{\Delta \delta}\left[-\frac{\Delta-\delta}{2}+(\xi-x_i)\right]$ & \\ [1.5ex]
       \hline\hline
     \end{tabular}
   \end{center}
   \caption{Charge deposition rule corresponding to the fractional particle shape \eqref{eq:S_Delta_h}.}
   \label{table:fractional_charge_deposition}
\end{table}

Another approach to obtaining particle shapes of arbitrary support
is based on the finite element method of discretizing a system of
equations \cite{becker_finite_1981}. In particular, in particle
algorithms based on a variational principle
\cite{lewis:1970:136,evstatiev_variational_2013}, the convolution
method of obtaining particle shapes emerges naturally \cite{evstatiev_variational_2013}
and the role of the above primary particle shape $S_0$, which
provides the connection to the grid, is taken by finite element basis functions.
For example, tent functions of width $2\Delta$ and height $1$
(linear Lagrange finite elements, which are basically the same as the linear spline
shape function in Table~\ref{table:particle_shapes} except with a different amplitude),
$\Psi_{i}(x) = \Psi(\xi-x_i)$, have the property that
at \emph{any} $x$ (including at grid points $x_i$)
\begin{equation}
  \sum_{i=1}^{N_{g}}\Psi_{i}(x)=1\,.
  \label{eq:FE_unary}
\end{equation}
Except for a factor $\Delta$, this is the sum rule that we require of particle shapes.
That is,
to obtain a particle shape $S(x)$ satisfying the sum rule \eqref{eq:S_sum_rule},
we perform the convolution 
$S(x)=\int_{0}^{1}\!K(y)\Psi(x-y)\,dy/\Delta$
with the finite element.
(Note that translation invariance is also satisfied).
The unit normalization follows from the finite element property
$\int_{0}^{1}\!(1/\Delta)\Psi_{i}(x)\,dx=1$, $i=1,2,\ldots,N_g$,
and the kernel normalization \eqref{eq:S_normalization-1}.
The shapes (top to bottom) from Table~\ref{table:particle_shapes}
may also be obtained by the finite element method from convolutions.

To summarize, relaxing the limitations associated with particle smoothness and
width allows one to devise particle shapes that are both computationally efficient and
suitable to take advantage of the BVO guidelines discussed in the previous section,
as well as assuring that the sum rule is satisfied.

\subsection{Density analysis in discrete variables}
\label{sec:discrete_density}

We consider a uniform grid in $x$ on $[0,1]$ with vertices at $x_{i}=\Delta i$,
$i=0,\cdots,N_{g}$ and grid spacing $\Delta=1/N_{g}$.
Returning to $\rho(x)=1$ for simplicity, we define the estimated density at
\emph{cell centers}, $x_{i+1/2}=(i+1/2)\Delta$, as
\[
\rho_{e,i+1/2}=\rho_{e}(x_{i+1/2}),\quad 0\le i\le N_{g}-1\,.
\]
Note that given a particle shape, e.g., from Table~\ref{table:particle_shapes},
charge deposition on cell centers amounts to simply substituting $x=\xi-x_{i+1/2}$
in $S(x)$.
Recall that for uniform density we have
$\langle\rho_{e,i+1/2}\rangle\equiv\left<\rho_{e}(x_{i+1/2})\right>=1$;
then the discrete approximation to $\int\rho_{e}(x)dx=1$ is 
\begin{equation}
  \sum_{i=0}^{N_{g}-1}\rho_{e,i+1/2}\Delta=\sum_{i,\mu} q_{\mu}S_{i+1/2}(\xi_{\mu})\Delta
  =\sum_\mu q_\mu\sum_i\Delta S(x_{i+1/2}-\xi_\mu)=1.
\label{eq:S_sum_rule_cell_centered}
\end{equation}
For the covariance matrix we have, from Eqs.~\eqref{eq:C_d_xy},
(\ref{eq:continuum-covariance}) 
\[
C_{i+1/2,j+1/2}\equiv C(x_{i+1/2},x_{j+1/2})=C_{d,i+1/2,j+1/2}-\frac{1}{N_{p}}
\]
\begin{equation}
=\frac{1}{N_{p}}\left[\int S(x_{i+1/2}-\xi)S(x_{j+1/2}-\xi)d\xi-1\right],\label{eq:discreteCij}
\end{equation}
the last equality due to the sum rule, Eq.~\eqref{eq:S_sum_rule}, and $\sum_\mu q_{\mu}=1$. 
The discrete analog of Eq.~(\ref{eq:intCdy=00003D00003D0}) is the condition
that the sum over each row (or column) of the covariance matrix is zero;
we have
\begin{align}
  \sum_{j}\Delta C_{i+1/2,j+1/2}
  = & \frac{1}{N_{p}}\left[\sum_{j}\Delta\int S(x_{i+1/2}-\xi)S(x_{j+1/2}-\xi)d\xi-1\right]\nonumber \\
  = & \frac{1}{N_{p}}\left[\Delta \sum_{j}\hat{S}(x_{i+1/2}-x_{j+1/2})-1\right]=0\,,
      \label{eq:Sum-rule-discreteCij}
\end{align}
where $\hat{S}$ is the convolution of $S$ with itself. Since $S$
satisfies the sum rule and assumptions \eqref{eq:S_normalization-1}--\eqref{eq:S_finite-1},
so does $\hat{S}$, therefore, the quantity
in Eq.~(\ref{eq:Sum-rule-discreteCij}) sums to zero and we obtain the analogous discrete result
as in Eq.~\eqref{eq:intCdy=00003D00003D0}. 
As in the continuous case, this identity says that the vector $u_i =1$ is associated with 
zero eigenvalue, implying that the covariance matrix is singular.
In the next section we will show exact calculations of these covariance
matrix elements for specific particle shapes.

The negative correlations of Eq.~(\ref{eq:continuum-covariance})
also appear in Eq.~(\ref{eq:discreteCij}). In order to understand
these correlations, let us compare a case in which we pick the number
$N_{i+1/2}$ of particles in the cells independently and identically
distributed (iid) from a Poisson distribution with parameter $\lambda=N_{ppc}=N_{p}\Delta$,
with mean $\lambda$ and variance $\lambda$. Here, $N_{ppc}$ is
the expected number of particles per cell. Again, recall that in this
discussion, we are assuming $\rho(x)=1$; thus the mean of $\rho_{e,i+1/2}=N_{i+1/2}/N_{ppc}$
is unity and its variance is $1/N_{ppc}$.
Indeed, from the iid assumption, the off-diagonal terms are zero and using
the property of the variance, $\mbox{Var}(\rho_{e,i+1/2})=\mbox{Var}(N_{1+1/2})/N_{ppc}^{2}$,
we obtain
\begin{equation}
C_{i+1/2,j+1/2}=\frac{\lambda}{N_{ppc}^{2}}\delta_{ij}=\frac{1}{N_{ppc}}\delta_{ij}.\label{eq:Poisson-result}
\end{equation}
Referring to Eqs.~(\ref{eq:K_definition}), (\ref{eq:discreteCij}),
for a kernel of width $h\sim\Delta$,
the diagonal (the variance) is $\sim1/\Delta N_{p}=1/N_{ppc}$,
comparable to the dominant part of the diagonal in Eqs.~\eqref{eq:discreteCij},
\eqref{eq:Sum-rule-discreteCij}. However, the negative correlations $-1/N_{p}$
in Eqs.~(\ref{eq:continuum-covariance}) and \eqref{eq:discreteCij}
are not contained in the Poisson model. These negative correlations
are traced to the fact that the total number of particles $N_{p}$
in Eqs.~\eqref{eq:continuum-covariance} and (\ref{eq:discreteCij}) is fixed.
In a particle code, the total number of particles can be assumed fixed at each time step.
(The particle number may also be constant throughout the simulation for
certain type of boundary conditions such as periodic, for example.)
The fixed number of particles is in contrast to the Poisson case in
which the \emph{expected} number of particles per cell is $N_{ppc}$
(expected total number of particles $=N_{p}$). Intuitively,
when the total number of particles is fixed,
if one cell has more than the expected number of particles $N_{ppc}$, other
cells must necessarily have fewer particles, leading to negative correlations.
(See also Ref.~\cite{ross_first_2006} where 
negative correlations between numbers of
particles in different cells were obtained working from the multinomial distribution.)
We will discuss in a later section the effect of these negative correlations
on the calculation of the electric field.

\subsection{Covariance $C_{i+1/2,j+1/2}$ examples}
\label{sec:discrete_covariance}

In this section we present covariance matrix calculations
with particle shapes from Table~\ref{table:particle_shapes}.
The simplest particle shape is the boxcar (top-hat) function.
For $i\ne j$ the overlap integral in Eq.~(\ref{eq:discreteCij}) is zero and we
find 
\[
C_{i+1/2,j+1/2}=-\frac{1}{N_{p}}\,\,\,(j\ne i)
\]
giving 
\begin{equation}
C_{i+1/2,j+1/2}=\frac{1}{N_{ppc}}\delta_{ij}-\frac{1}{N_{p}}.\label{eq:NGP_covariance}
\end{equation}
The first term in Eq.~\eqref{eq:NGP_covariance} is equal to the
value in Eq.~(\ref{eq:Poisson-result}), and the second term is recognized
as the negative term of Eq.~(\ref{eq:discreteCij}). We conclude
indeed that the $-1/N_{p}$ term is due to the fact that the total
number of particles is fixed.
The condition \eqref{eq:Sum-rule-discreteCij} is obviously satisfied
in this example.

For the linear particle shape we find
\begin{equation}
C_{i+1/2,j+1/2}=\left\{ \begin{tabular}{ll}
                          \ensuremath{\frac{1}{N_{p}}\left[\Delta\int S(\xi)^{2}d\xi-1\right]
                          =\frac{1}{N_{p}}\left[\frac{2}{3\Delta}-1\right]=\frac{2}{3N_{ppc}}-\frac{1}{N_{p}}\,\,\,}  &  \ensuremath{(j=i)}, \\
                          \rule{0pt}{4ex} \ensuremath{\frac{1}{N_{p}}\left[\Delta\int S(\xi-1)S(\xi)d\xi-1\right]
                          =\frac{1}{6N_{ppc}}-\frac{1}{N_{p}}\,\,\,}  &  \ensuremath{(j = i\pm 1)}, \\
 \rule{0pt}{4ex} \ensuremath{-\frac{1}{N_{p}}\,\,\,}  &  \mbox{otherwise}. 
\end{tabular}\right.\label{eq:linear_covariance}
\end{equation}
Note that the condition \eqref{eq:Sum-rule-discreteCij} holds for this example as well.
This condition highlights the importance of the $-1/N_{p}$ correlations
and shows how the variances ($i=j$ terms) decrease as the particle
width increases.

For the quadratic particle shape we have
\begin{equation}
C_{i+1/2,j+1/2}=\left\{ \begin{tabular}{ll}
 \ensuremath{\frac{11}{20N_{ppc}}-\frac{1}{N_{p}}\,\,\,}  &  \ensuremath{(j=i)} \\
 \rule{0pt}{3ex} \ensuremath{\frac{13}{60N_{ppc}}-\frac{1}{N_{p}}\,\,\,}  &  \ensuremath{(j=i\pm1)} \\
 \rule{0pt}{3ex} \ensuremath{\frac{1}{120N_{ppc}}-\frac{1}{N_{p}}\,\,\,}  &  \ensuremath{(j=i\pm2)} \\
 \rule{0pt}{3ex} \ensuremath{-\frac{1}{N_{p}}\,\,\,}  &  \mbox{otherwise}. 
\end{tabular}\right.\label{eq:quadratic_covariance}
\end{equation}
The property \eqref{eq:Sum-rule-discreteCij}
is again clearly seen for all these cases.
Note again that the variances, i.e.~the $i=j$ terms, decrease further as the particle
widths increase, with their values distributed to more neighboring
bands, with the negative term $-1/N_p$ on diagonal as well as off-diagonal terms.

\subsection{Discretized electric field correlations}
\label{sec:discrete_Efield}

In this section we analyze the statistical properties of the electric
field on a grid, again assuming uniform density, $\rho(x)=1$. The continuous
form of these relations was presented in Sec.~\ref{sec:Electric-field-computation}.
In the previous section, the density was assumed to reside at cell
centers $\rho_{e,i+1/2}$, i.e., at $x_{i+1/2}$. If the density and
the electric field are taken to be staggered, with $E_{i}$ residing
at vertices $x_{i}$ for $i=0,\cdots,N_{g}$, to obtain a second order
accurate differencing scheme,\footnote{This discretization can also be derived using finite elements.}
we write 
\begin{equation}
E_{i+1}-E_{i}=\Delta\rho_{q,i+1/2},\label{eq:Gauss-difference}
\end{equation}
where $\rho_{q,i+1/2}=1-\rho_{e,i+1/2}$ is the charge density and, as before,
assuming uniform and immobile ions of unit (neutralizing) density.
This relation leads to $E_{i}=E_{0}+E_{1,i}$, where 
\begin{equation}
E_{1,i}=\Delta\sum_{j=0}^{i-1}\rho_{q,j+1/2}.\label{eq:Gauss-sum}
\end{equation}
The condition $E_{0}=E_{N_{g}}$ ($x_{0}=0,\,\,x_{N_{g}}=1$) follows from 
\begin{equation}
\sum_{j=0}^{N_{g}-1}\rho_{q,i+1/2}=0\label{eq:sum-of-rho=00003D00003D0}
\end{equation}
and Eq.~\eqref{eq:S_sum_rule_cell_centered}. (It is instructive
to revisit the derivation of the discrete property \eqref{eq:Sum-rule-discreteCij}:
this result can be obtained directly by writing $\sum_{j}\langle\tilde{\rho}_{e,i+1/2}\tilde{\rho}_{e,j+1/2}\rangle$
$=\langle\tilde{\rho}_{q,i+1/2}\sum_{j}\tilde{\rho}_{q,j+1/2}\rangle$,
which is seen to vanish by Eq.~(\ref{eq:sum-of-rho=00003D00003D0}).)
The condition in Eq.~\eqref{eq:ZeroPotential}
of having zero applied potential across a period takes the discrete
form 
\begin{equation}
  \sum_{i=0}^{N_{g}-1}\Delta E_{i}=\Delta N_{g}E_{0}+\sum_{i=1}^{N_{g}-1}E_{1,i}=0.
  \label{eq:Zero-potential}
\end{equation}
This is a condition on $E_{0}$; we find 
\begin{align*}
E_{0}= & -\Delta^{2}\sum_{i=0}^{N_{g}-1}\sum_{j=0}^{i-1}\rho_{q,j+1/2}\\
= & -\Delta^{2}\sum_{j=0}^{N_{g}-2}(N_{g}-j-1)\rho_{q,j+1/2}.
\end{align*}

\begin{figure}[t!]
  \begin{center}
    \includegraphics[width=0.7\textwidth]{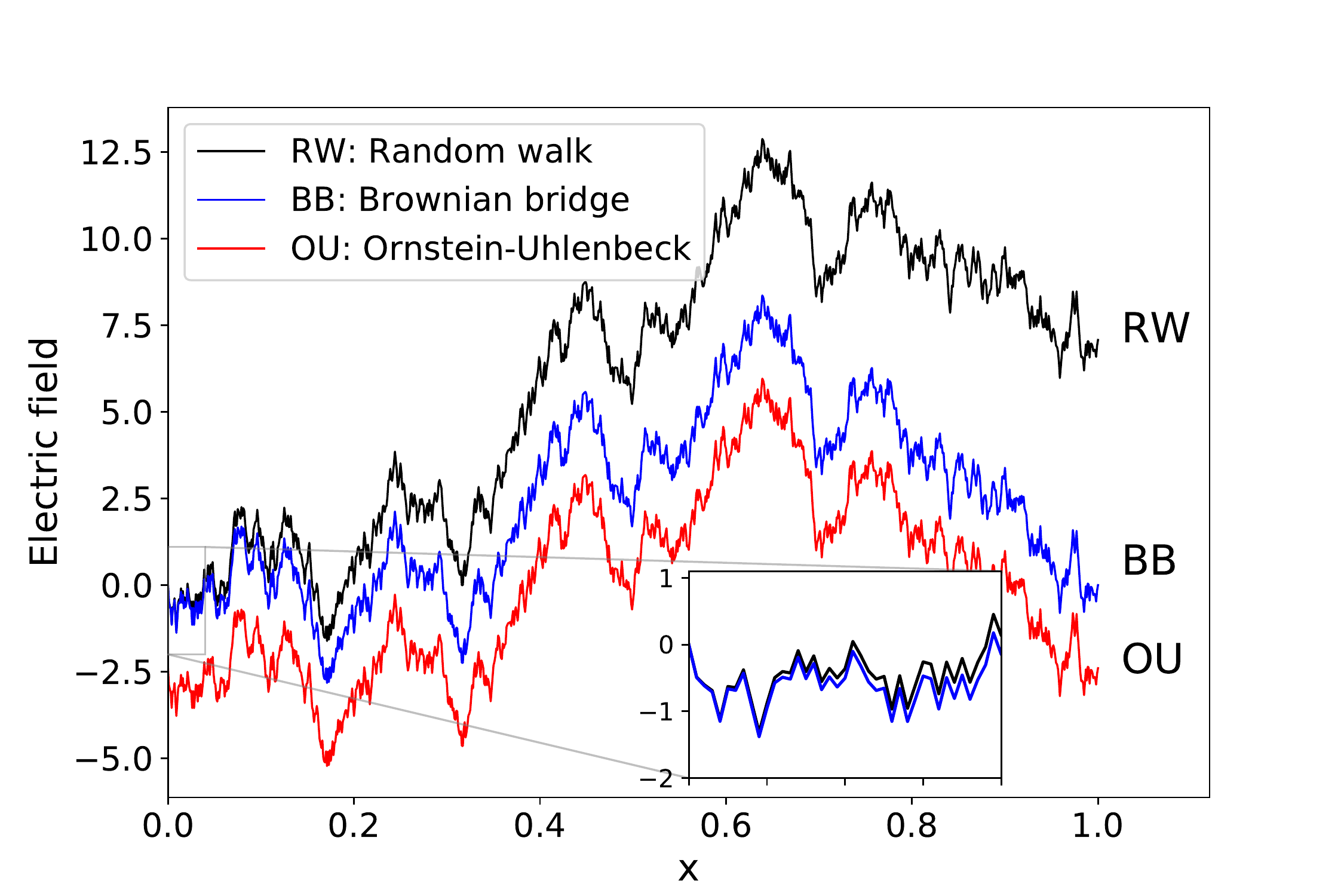}
  \end{center}
  \caption{Illustration of three types of random behavior of the electric field
    depending on the boundary conditions, as discussed in Appendix~\ref{app:brownian_bridge}.
    Both RW and BB have a starting point $E(0)=0$ (see inset);
    the BB case has additionally
    $E(1)=0$, and the OU case is shifted down, satisfying the zero potential difference
    condition. }
  \label{fig:E_random_walk}
\end{figure}

As in Sec.~\ref{sec:Electric-field-computation}, the terms in Eq.~(\ref{eq:Gauss-sum})
lead to four distinct terms in the covariance matrix for the noise
in the electric field, $\langle\tilde{E}_{i}\tilde{E}_{j}\rangle=C_{00}^{E}+C_{10,i}^{E}+C_{01,j}^{E}+C_{11,ij}^{E}$,
where 
\begin{equation}
C_{00}^{E}=\langle\tilde{E}_{0}^{2}\rangle,\label{eq:C00-more}
\end{equation}
\begin{equation}
C_{10,i}^{E}=\langle\tilde{E}_{1,i}\tilde{E}_{0}\rangle,\label{eq:C10-C01}
\end{equation}
similarly for $C_{01,j}^{E}$, and 
\begin{equation}
C_{11,ij}^{E}=\langle\tilde{E}_{1,i}\tilde{E}_{1,j}\rangle.\label{eq:C11}
\end{equation}
For the first of these we find 
\begin{equation}
C_{00}^{E}=\Delta^{4}\sum_{j=0}^{N_{g}-2}(N_{g}-j-1)\sum_{k=0}^{N_{g}-2}(N_{g}-k-1)C_{j+1/2,k+1/2},\label{eq:C00-more}
\end{equation}
where $C_{j+1/2,k+1/2}=\langle\tilde{\rho}_{q,j+1/2}\tilde{\rho}_{q,k+1/2}\rangle.$
For the next term (and similarly for $C_{01,j}^{E}$) we have 
\begin{equation}
C_{10,i}^{E}=-\Delta^{3}\sum_{j=0}^{i-1}\sum_{k=0}^{N_{g}-2}(N_{g}-k-1)C_{j+1/2,k+1/2}.\label{eq:C01-C10-more}
\end{equation}
Finally, the last term is 
\begin{equation}
C_{11,ij}^{E}=\Delta^{2}\sum_{k=0}^{i-1}\sum_{l=0}^{j-1}C_{k+1/2,l+1/2}.\label{eq:C11-more}
\end{equation}
For this covariance matrix we find 
\[
  C_{11,ij}^{E}=\frac{1}{N_{p}}\left(\Delta\min(i,j)-\Delta^{2}ij\right)
  =\frac{1}{N_{p}}\left(\min(x_{i},x_{j})-x_{i}x_{j}\right).
\]
The first term is the value for the Poisson case and the second is
the Brownian bridge contribution from the constant correlations $-1/N_{p}$
in Eq.~(\ref{eq:NGP_covariance}). The remaining terms arise from
the Ornstein-Uhlenbeck bridge; we have 
\[
  C_{10,i}^{E}=\frac{1}{N_{p}}\left(-\Delta^{2}(N_{g}-1)i
    +\frac{\Delta^{2}i(i-1)}{2}+\frac{\Delta^{3}N_{g}(N_{g}-1)i}{2}\right)
\]
\[
\rightarrow\frac{1}{N_{p}}\left(\frac{x_{i}^{2}}{2}-\frac{x_{i}}{2}\right),
\]
the latter limit as $\Delta\rightarrow0$ with $\Delta N_g$=1. The quantity $C_{01,j}^{E}$
is computed similarly and 
\[
  C_{00}^{E}=\frac{1}{N_{p}}\left(\Delta^{3}\frac{(N_{g}-1)N_{g}(2N_{g}-1)}{6}
    -\Delta^{4}\frac{N_{g}^{2}(N_{g}-1)^{2}}{4}\right)
\]
\[
\rightarrow\frac{1}{12N_{p}}
\]
and in the same limit, $\Delta \rightarrow 0$.
We see that the limiting case is in agreement with Eq.~\eqref{eq:Full-CE(x,y)}.

The three different types of random behavior of the electric field, depending on the
boundary conditions, is illustrated in Fig.~\ref{fig:E_random_walk}. The random walk curve is a
Brownian motion with $E(0)=0$ without imposing any boundary condition at $x=1$, the Brownian bridge curve
reflects the extra boundary condition $E(0)=E(1)=0$, and the Ornstein-Uhlenbeck curve
satisfies both $E(0)=E(1)=0$ and the zero potential difference condition.
The latter means that if $E(x)$ makes an excursion into the positive half plane,
it must do so in negative half plane as well so that $E(x)$ integrates to zero -- see
the discussion in terms of position, velocity, and acceleration
in Appendix~\ref{app:brownian_bridge}.

In the next section we verify numerically our theoretical conclusions.

\section{Numerical results}
\label{sec:numerical}

Considerations dictating the choice of a charge deposition rule in a particle
algorithm were discussed in Sec.~\ref{sec:sum_rule}, where two important factors were
identified -- smoothness and width.
In sections \ref{sec:stat_uniform} and \ref{sec:stat_nonuniform}
we have found the particle width to be more relevant to our discussion:
it reduces the variance for uniform density and minimizes the error in
a non-uniform density via the BVO process, independent of grid resolution. For discretization 
on a grid, we have also emphasized the importance of obeying the sum rule.

One strategy for applying our theory in practice is to
use particles that have width closest to the optimal,
i.e., $h=i\Delta + \delta \approx \Hopt$ with $\Hopt$ given by \eqref{eq:h-opt},
$i$ an integer number, and $\Delta$ chosen according to a desired
grid resolution. One may further decide on a particle that is an integer
number of cells wide or add the additional correction $\delta$.
Using particles with exact width equal to $\Hopt$ is obviously not
possible in general simulations where the exact density and its gradients are unknown
and may vary in time. However, the estimated quantities can be used as a guideline
or a lower resolution simulation may be done to probe for these
and other properties of a physical system.
As discussed previously, spline functions of order $i>4$ are
probably not the most efficient choice and custom particles
may be better suited. If one uses high grid resolution (small $\Delta$),
one may be able to approximate $\Hopt$ sufficiently well with an integer number
of cells, $h=i\Delta\approx \Hopt$;
However, one advantage of using a fractional width shape with a
correction $\delta$ is that the same
charge deposition can be used and adjusted ``in real time,'' depending
on the values of $\rho(x)$ and $\rho''(x)$. This provides
``fine tuning'' ability, which may be preferable to changing the type/width of
particle in the course of a simulation and may help to avoid
introducing undesired effects or difficulties.

In the following sections, our focus will be on theory comparison and verification,
which is why we will not be concerned with the requirement of grid resolution.
Instead, we will use the grid spacing to adjust the width of
the \emph{same} particle shape, thus applying the scaling
transform $K(x)=K_f(x/h)/h$. When using the scaling method,
we will not be dealing with fractional width particles; therefore, to change
the width of a given particle, we will vary the grid spacing by integer
numbers: for example, when $N_{g}=15$, the width of the three-cell-wide quadratic
spline equals $h=3\Delta=3/15=0.2$, for $N_{g}=30$ its width equals
$h=3/30=0.1$, etc. Recall that we work in the domain $[0,1]$;
a range of $N_{g}\simeq 15\ldots 50$ will prove to provide a sufficient
range of particle widths. In a separate set of simulations, we
present results on a fixed grid but varying the fractional width
of the  particle defined in \eqref{eq:S_Delta_h}.

\subsection{Covariance matrix in uniform density}
\label{sec:numerical_correlations}

We present as a first example numerical computations of the density
covariance matrix with uniform true density $\rho(x)=1$ on a uniform
periodic grid on $[0,1]$ 
and we use the \emph{linear} particle shape from Table~\ref{table:particle_shapes}
for our charge deposition.
Recall that the covariance matrix for this case is given by Eq.~\eqref{eq:linear_covariance};
however, more convenient quantities to test are the products 
\begin{align}
\widetilde{C}_{i+\frac{1}{2},i+\frac{1}{2}}\equiv C_{i+\frac{1}{2},i+\frac{1}{2}}\times N_{ppc} & =\frac{2}{3}-\Delta\,,\\
\widetilde{C}_{i+\frac{1}{2},i+\frac{1}{2}\pm1}\equiv C_{i+\frac{1}{2},i+\frac{1}{2}\pm1}\times N_{ppc} & =\frac{1}{6}-\Delta
\end{align}
since they are independent of $N_{ppc}$ and $N_{p}$.

The theory was developed in the limit of infinite number of samples
by integrating over a continuous density distribution but
clearly, numerically we can only use a finite number of samples in averages.
The present results aim to verify the developed theory as well as to inform us of the
number of samples needed in simulations in the next section.
Because in this section we deal with uniform density, the correlations
are expected to be the same for every grid point.
Using this fact, we can obtain better statistics by averaging correlations
over the whole grid, i.e., 
$\bar{C}_{i+\frac{1}{2},i+\frac{1}{2}}=(1/N_{g})\sum_{1}^{N_{g}}C_{i+\frac{1}{2},i+\frac{1}{2}}$
for \emph{every} $i=1\ldots N_g$.
Therefore, for a number of $M_{\rm local}$ samples, such averaging amounts to
performing $M = N_{g}\times M_{\rm local}$ local cell samples; to the end of this
section we omit the over-bar.
Each particle sample is drawn from a
uniform distribution by drawing $N_{p}$ random numbers $R\in[0,1]$
and setting particle positions $\xi_{\mu}=R$, $\mu=1,2,\ldots N_{p}$.

Simulation results on a fixed grid with $N_{g}=25$ ($\Delta=0.04$) are listed in
Table~\ref{table:uniform_density}. Theory predicts
$\widetilde{C}_{i+\frac{1}{2},i+\frac{1}{2}}=0.666\ldots-0.04=0.6266\ldots$
and $\widetilde{C}_{i+\frac{1}{2},i+\frac{1}{2}\pm1}=0.1666\ldots-0.04=0.1266\ldots$.
\begin{table}[h]
  \footnotesize
\centering
\begin{tabular}{c|c|c|c|c|c}
  \hline\rule{0pt}{3ex}
  \multirow{2}{*}{$N_{p}$}  & \multirow{2}{*}{$M$} & \multicolumn{2}{c|} {$\widetilde{C}_{i+\frac{1}{2},i+\frac{1}{2}}$}
  & \multicolumn{2}{c}{$\widetilde{C}_{i+\frac{1}{2},i+\frac{1}{2}\pm1}$} \\
\cline{3-6}
 &  & theoretical  & numerical  & theoretical  & numerical \\
\hline 
\hline 
250  & $2.5\times 10^6$  & 0.6266\ldots{} & 0.6269  & 0.1266\ldots{} & 0.1267 \\
\hline 
2500  & $2.5\times 10^5$  &  & 0.6256  &  & 0.1251 \\
\hline 
25,000  & $2.5\times 10^4$  &  & 0.6208  &  & 0.1252 \\
\hline \hline
\end{tabular}\caption{Simulation results for uniform density distribution.
  The theoretical values have the fixed quantity $\Delta$ subtracted.
    In the numerical values, this quantity is not subtracted explicitly.}
\label{table:uniform_density} 
\end{table}
We have also intentionally taken the product $N_{p}\times M=\mbox{const.}$
in order to examine the role of number of samples versus number of
particles. The numerical values in the table agree with the theoretical
values of the correlations and are most accurate (to $3$ significant
figures) when using the largest number of samples $M=2.5\times10^{6}$ --
the first row in Table~\ref{table:uniform_density}.
Using a larger number of particles and smaller number of samples shows
good agreement as well, albeit with a
somewhat larger error in the third significant figure.
We conclude that a number of samples of order $10^{6}$ should be
sufficient for comparison with theory, with three significant
figures and a small error bar in the third significant figure.

\subsection{Bias-variance optimization in non-uniform density}
\label{sec:numerical_bias_variance}

This set of simulations aims to compare numerical results and theory for
the \emph{local} error in a \emph{non-uniform} density, $\rho(x)\ne\mbox{const.}$
We compare the minimum error and optimal particle width for four different
shapes, all having the same support $h=3\Delta$: the boxcar shape scaled to $3\Delta$,
the quadratic spline and the trapezoidal shape from Table~\ref{table:particle_shapes},
and the Epanechnikov kernel scaled to $3\Delta$,
\begin{equation}
  K_{E}(x)=\frac{1}{\Delta}\left\{
    \begin{tabular}{ll}
      \ensuremath{\frac{1}{2}\left(1-\frac{4}{9}\left(\frac{x}{\Delta}\right)^{2}\right),}
      &  \ensuremath{\left|\frac{x}{\Delta}\right|\le \frac{3}{2}}\\[1.25ex]
      \ensuremath{0}  &  otherwise\,. 
    \end{tabular}\right.
  \label{eq:S_Epanechnikov}
\end{equation}
These four shapes also have different smoothness, which is another
point of comparison. We note that the wider boxcar shape satisfies the sum rule
  just as the $\Delta$-wide NGP shape does but deposits constant amount of charge
  on three grid points instead of one.
Unlike the first three shapes, the
Epanechnikov kernel \eqref{eq:S_Epanechnikov} does not satisfy the
sum rule and therefore does not conserve charge on the grid, i.e., is not a
particle shape in the sense of Sec.~\ref{sec:sum_rule}. For this reason,
the Epanechnikov kernel is not recommended for practical applications;
however, for our purpose of statistical calculations, charge conservation
is not required. The trapezoidal shape (which does satisfy the sum rule)
is also unusual and since we have not tested it in full particle simulations,
we do not recommend it for practical use at this point.
The purpose of present
comparisons is to (i) verify the theoretical prediction that
the minimal error depends primarily on the
particle width and not significantly on the specific particle shape (and smoothness);
and (ii) to verify the general dependence of $\Qmin$ and $\Hopt$
on the particle shape.\footnote{We remind the reader that $Q$ is the
  mean-square error and the actual error is $\sqrt{Q}$.}

For all simulations in this section the true density distribution
is given by the periodic function
\begin{equation} \rho(x) = 1 + a\cos(2\pi m x)\,,
  \label{eq:cos_density}
\end{equation}
where $a<1$ is a constant amplitude and $m$ is an integer mode
number. Since \eqref{eq:cos_density} satisfies
$\int_{0}^{1}\rho(x)\,dx=1$ and $\rho(x)\ge 0$ for $x\in[0,1]$,
it is a probability density distribution.
For all simulations in this section we have chosen $a=1/2$, and $m=2$.
We show results for two locations: $x=1/2$, where $\rho(1/2)=3/2$ and $\rho''(1/2)=-8\pi^{2}$,
and $x=1/3$, where $\rho(1/3)=3/4$ and $\rho''(1/3)=-4\pi^{2}$.
The number of samples is $M=10^{6}$ and the number of particles is
$N_{p}=10^{4}$, unless otherwise stated.
Drawing a sample from the density \eqref{eq:cos_density} was done by
by the transformation method \cite{Press_1986}, by which the cumulative distribution
function of $\rho(x)$ is inverted; this process is repeated
$\mu=1\ldots N_p$ times for the particles in the sample.
The whole process is repeated for a total of $M$ samples.
We note that by connecting particle widths to integer
number of cells, we do not reach the absolute theoretical minimum
$\Hopt$ since $N_{g}$ in a domain of fixed size can only be
an integer number and $\Delta=1/N_{g}$ can only take discrete values.

\begin{figure}[!h]
\begin{centering}
\includegraphics[width=0.7\textwidth]{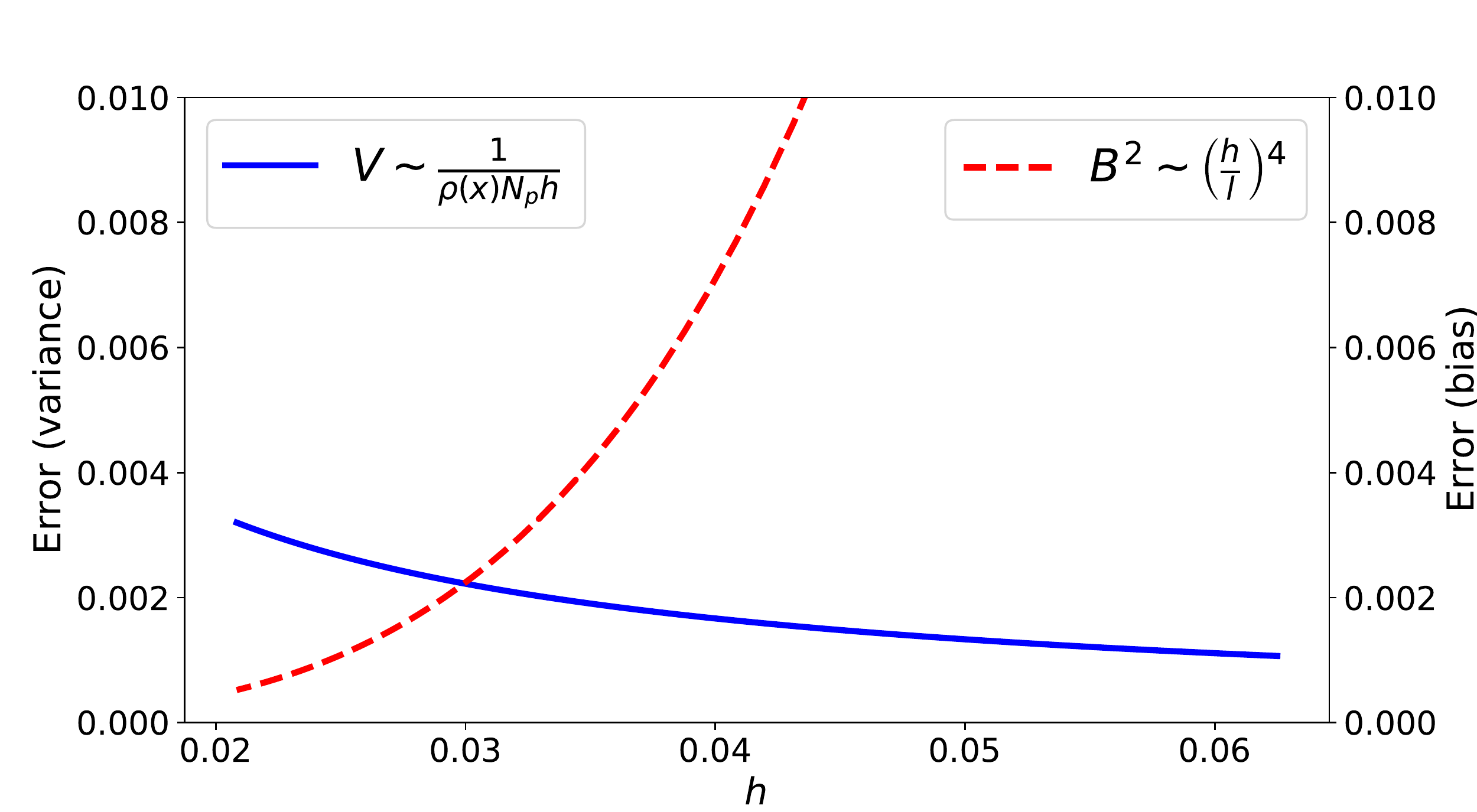} 
\par\end{centering}
\caption{ Qualitative comparison of the terms representing the bias and variance
    errors in Eq.~\eqref{eq:BV_balance}.
}
\label{fig:BVO_balance} 
\end{figure}

Before discussing the numerical simulations, let us look
qualitatively at the error minimum for the density \eqref{eq:cos_density}, based
on the order of magnitude estimates \eqref{eq:BV_balance}.
We assume the kernel width is a multiple of the grid spacing and 
for simplicity take $h=\Delta$, i.e., we estimate the density with a
boxcar shape. 
Using the scaling transformation \eqref{eq:K_definition},
we vary the grid in the range $N_g\in [16,48]$, which corresponds to
kernel widths in the range $h\in[0.02083,0.0625]$. 
We need to compare $h$ with, the density gradient scale length, which for
our case of Eq.~\eqref{eq:cos_density} at $x=1/2$ gives $l = \sqrt{1.5/8\pi^2}\simeq 0.138$.
The qualitative bias and variance error curves are illustrated in
Fig.~\ref{fig:BVO_balance}, showing that we should expect a minimum in this
range of simulation parameters. Notice that the lowest value of $h$ corresponds to
$N_g=48$ and is an acceptable grid resolution for our cosine density profile, having
$24$ grid points per period; it does not, however, correspond to $\Hopt$, 
the error being about $50\%$ above the actual minimum.
If we were to set up a grid corresponding to $\Hopt$ (using the same particle, with $h=\Delta$),
it  would correspond to $N_g\simeq 32$ and we would be somewhat under-resolving the cosine period
with only about $16$ grid points per period. We shall not further discuss this
issue but emphasize that grid resolution is a factor that
must be chosen based on grid truncation errors and \emph{independently} from the
particle shape, whose width is chosen to minimize $Q=V+B^2$,
where the variance $V$ is a measure of the statistical error.
Finally, the quantity $N_h$ from Eq.~\eqref{eq:BV_balance_Nh}
is equivalent to the number of particles per cell, $N_{ppc}$, in the discretized system.
For a total of $N_p=10^4$ particles, we find $N_{ppc}\in[312,937]$ for $N_g\in[16,48]$,
and $N_{ppc}\simeq 450$ at the intersection of the two curves,
which is close to the minimum of $Q=V+B^2$. Note that
in the intersection region the bias curve depends more strongly
on $h$ than the variance curve, suggesting that
a more efficient way to ``locate'' (and follow) the minimum of the error curve is by
adjusting $h$ rather than by adjusting the number of particles in a simulation.
One should keep in mind that the exact numerical parameters discussed above may
differ from the exact figures by a factor of a few,
coming from the shape coefficients $C_1$ and $C_2$, which were not included
in the estimates \eqref{eq:BV_balance} and \eqref{eq:BV_balance_Nh}.

\begin{figure}[!h]
\begin{centering}
\includegraphics[width=0.7\textwidth]{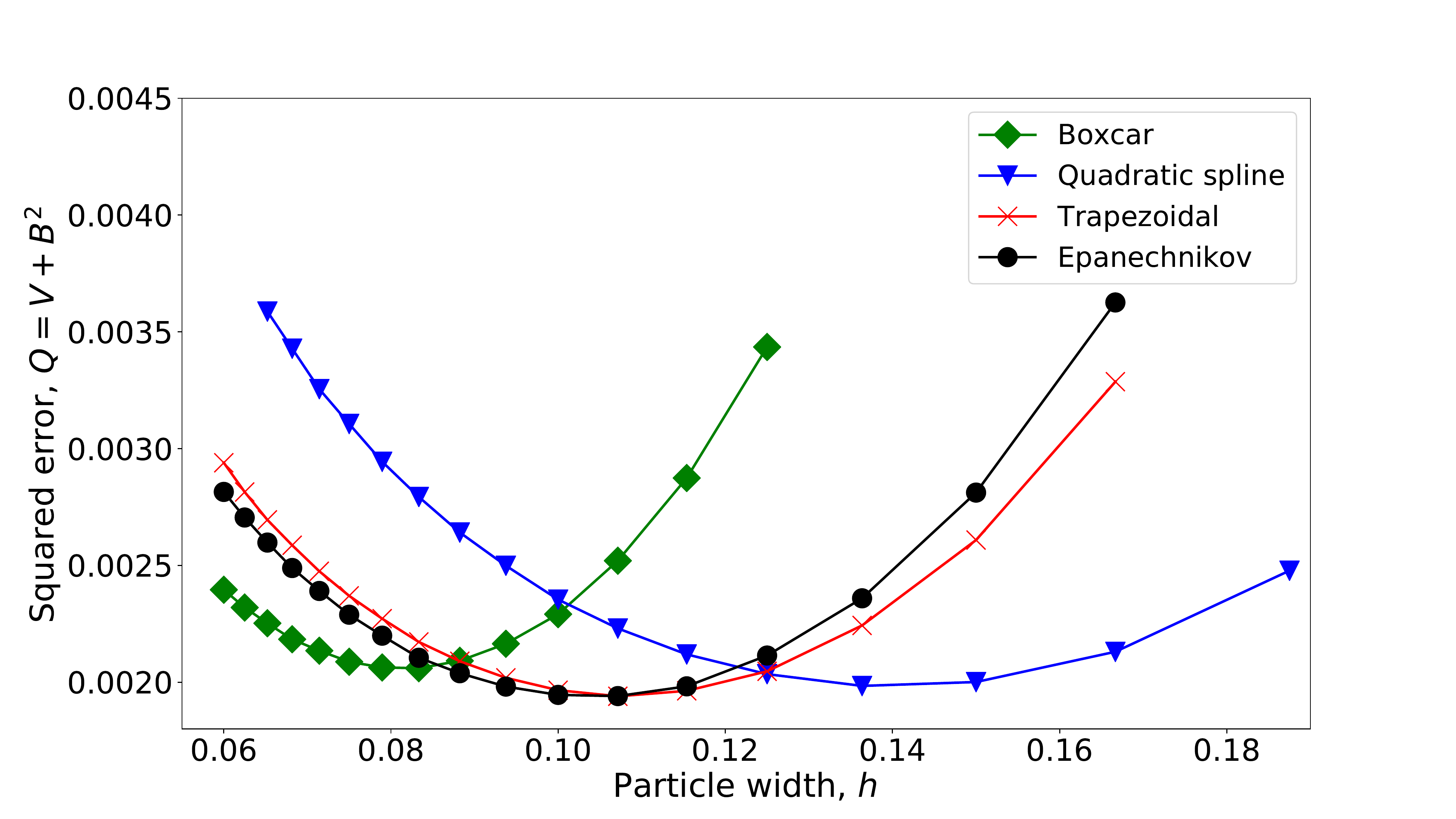} 
\par\end{centering}
\caption{Local bias-variance optimization comparison of four different shapes at
  $x=1/2$, from numerical computations.
  Although the value of $\Hopt$ changes noticeably between the different shapes,
  the value of $\Qmin$ depends little on the shape of the particle,
  in agreement with the results summarized in Table~\ref{table:Kernel_C_1_C_2}.}
\label{fig:BVO_comparison} 
\end{figure}

Numerical results for the above setup are shown in Fig.~\ref{fig:BVO_comparison}
at the location $x=1/2$. In this simulation we use the scaling
transform, Eq.~\eqref{eq:K_definition}, which is accomplished by
changing the grid spacing $\Delta$ (or $N_g$) and thus the particle
width $h=3\Delta = 3/N_g$. In the figure each
point corresponds to a different $N_g$ but we remind the reader that the
grid resolution is irrelevant to BVO calculations and thus
we should not expect interference with the results for $\Hopt$ and $\Qmin$.
As predicted by theory, a minimum of the error is achieved at some value $\Hopt$.
Further, the value of the
minimal error changes very little between the four different shapes.
The value of $\Hopt$ is seen to vary, with the 
exception of the trapezoidal shape and the Epanechnikov kernel,
which show similar values for both $\Hopt$ and $\Qmin$.
To understand the similarities and differences better, we use the values of
the constants $C_{1}$ and $C_{2}$ from Table~\ref{table:Kernel_C_1_C_2}
to calculate the theoretical $\Qmin$ and $\Hopt$ from
Eqs.~\eqref{eq:h-opt} and \eqref{eq:Qmin}. The results in
Table~\ref{table:h0_Qmin} confirm our observations
and also show good agreement between theoretical and numerical
calculations.

One important point concerning comparisons between numerical and theoretical 
results should be made. The theoretical calculations involved certain approximations
such as termination of the Taylor expansion in Eq.~\eqref{eq:<rhoe>-with-bias}
and neglecting terms $\sim 1/N_p$ in Eq.~\eqref{eq:1st-part-of-Q1=00003D00003Dagain}.
In contrast, the numerical results are in this sense ``exact'' since
no approximations are made; the only inaccuracy in the numerical
results stems from the finite number of samples used in the
averages. It is plausible
that certain density profiles may require keeping
some of the other neglected terms to better describe the optimal width and error. This
situation may occur, for example, when at isolated locations, $x_{0}$,
the second derivative of the density vanishes, $\rho''(x_{0})=0$.
Such occurrences are exceptions rather than the general situation
and of course, the averaged theory, Eqs.~\eqref{eq:h-opt-1}, \eqref{eq:MISE-Qmin},
remains unaffected. For the density profile \eqref{eq:cos_density} we will 
see below that at the specified above spatial locations, including one extra term in
the theoretical calculations further improves the agreement by a few percent.

\begin{figure}[!h]
\begin{centering}
\includegraphics[width=0.7\textwidth]{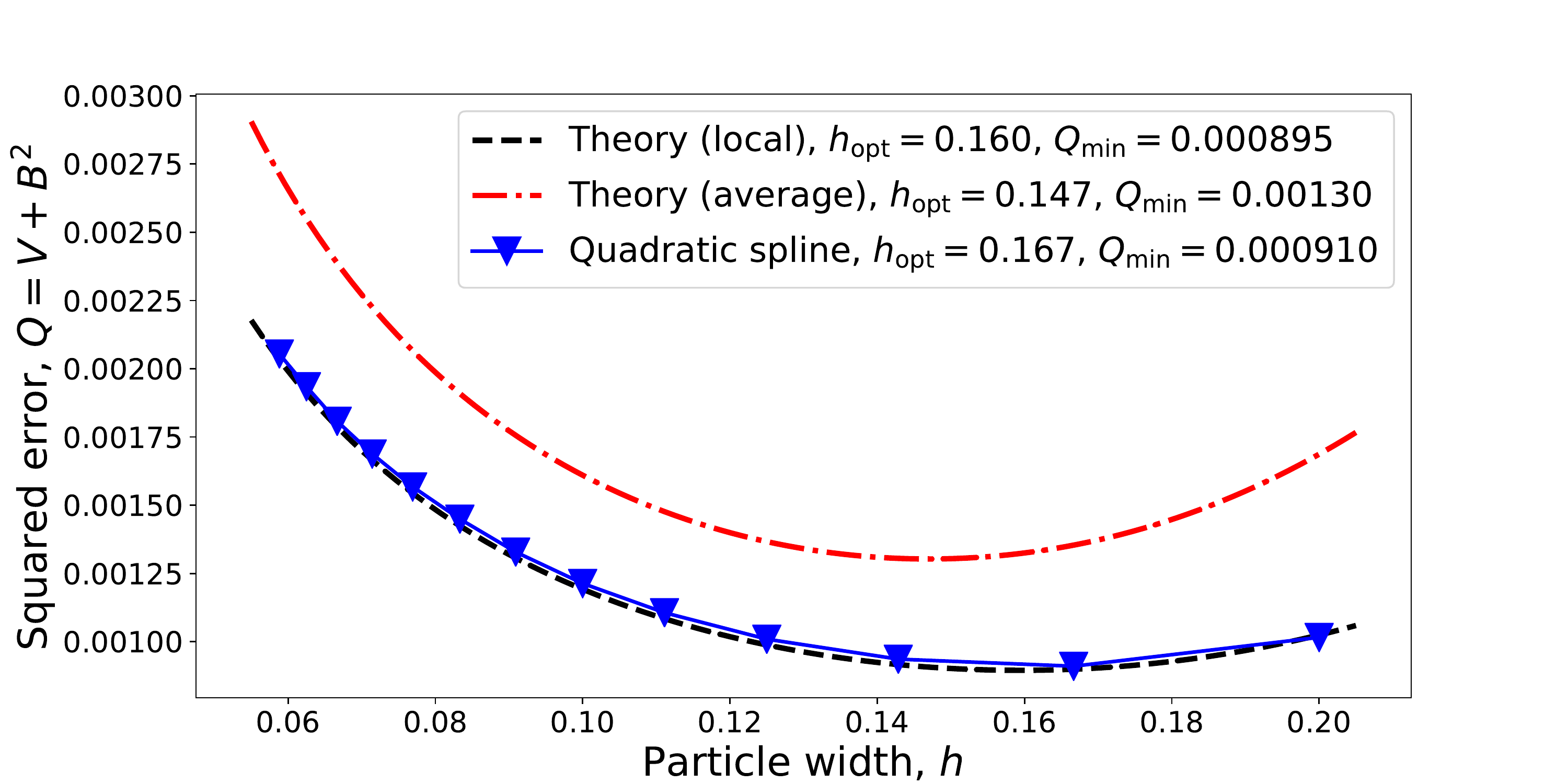}\\
 \includegraphics[width=0.7\textwidth]{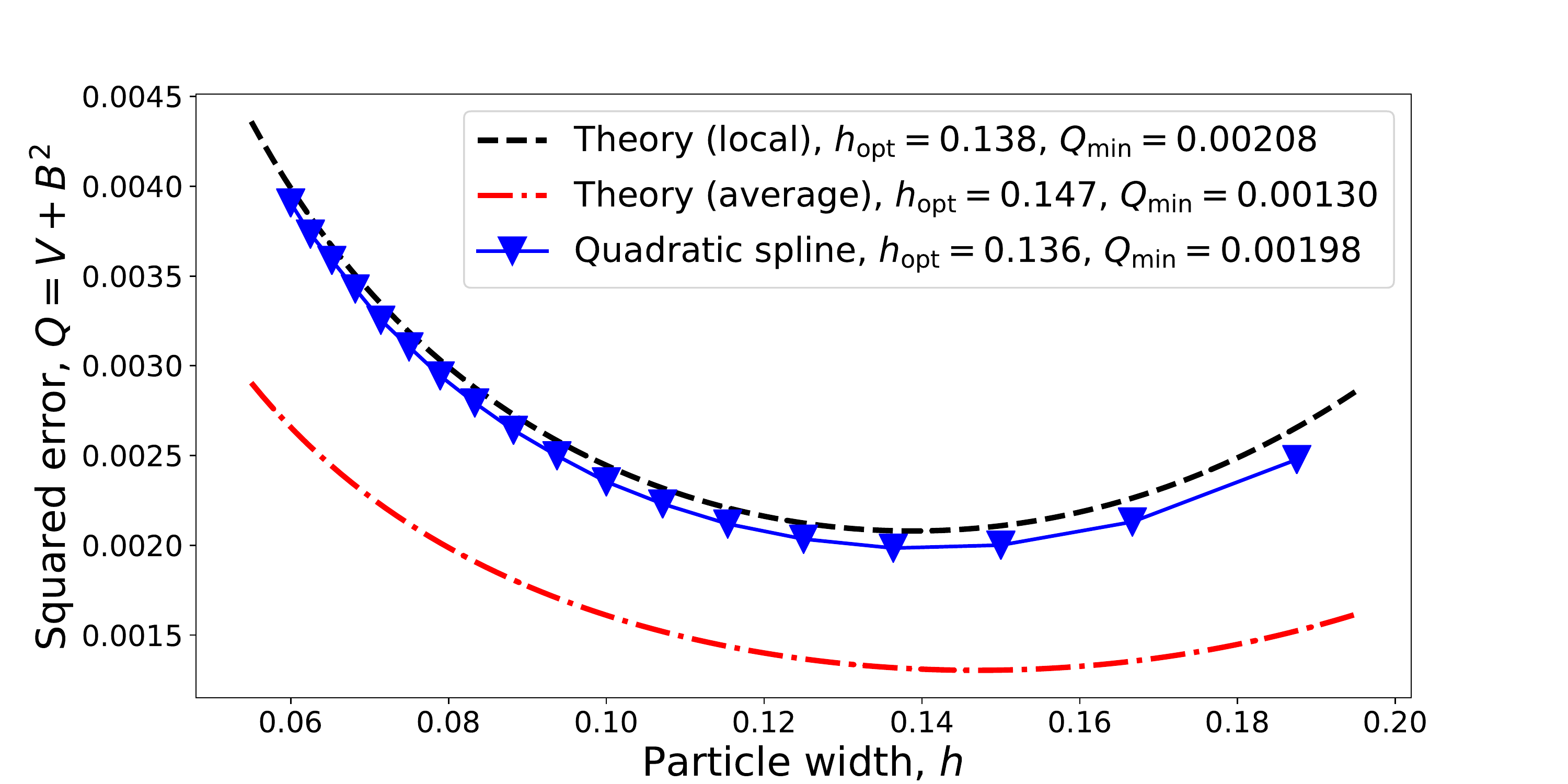} 
\par\end{centering}
\caption{Bias-variance optimization comparison between exact local theory, averaged theory,
and simulations for the quadratic spline particle shape. Top panel:
$x=1/3$. Bottom panel: $x=1/2$.}
\label{fig:BVO_comparison_quadr_spline} 
\end{figure}

In Figure~\ref{fig:BVO_comparison_quadr_spline} we show
comparison between local theory, Eqs.~\eqref{eq:h-opt}, \eqref{eq:Qmin},
averaged theory, Eqs.~\eqref{eq:h-opt-1}, \eqref{eq:MISE-Qmin},
and simulations for the two locations, $x=1/3$ (top panel) and $x=1/2$
(bottom panel), for the \emph{same} quadratic spline particle shape
(cf. Table~\ref{table:particle_shapes}).
The lines labeled ``Theory (local)'' are plots of formula \eqref{eq:Q=00003D00003DV+B^2};
the lines labeled ``Theory (average)'' are plots of the integrated (from $0$ to $1$)
Eq.~\eqref{eq:Q=00003D00003DV+B^2}.
Both theory and averaged theory use the values
of $C_{1}$ and $C_{2}$ from Table~\ref{table:Kernel_C_1_C_2}. We see that
the averaged theory does not change between the top and bottom panels
since $\Hopt$ and $\Qmin$ do not depend on $x$. (The apparent
difference is due to the slightly different plotting range.)
In order to obtain even better
agreement between local theory and simulations, we have
included the small correction $-\rho(x)^2/N_{p}$ in
Eq.~\eqref{eq:1st-part-of-Q1=00003D00003Dagain},
which has been neglected so far; that yields an improvement
of about $5\%$.  Including that term also explains the slight
difference in the theoretical values of $\Hopt$ and $\Qmin$ that the
discerning eye would observe in the legend of Fig.~\ref{fig:BVO_comparison_quadr_spline}
(bottom panel) versus the table values in Table~\ref{table:h0_Qmin} (Quadr. spline).

Lastly, we present simulations of bias-variance optimization
with the fractional width particle shape, Eq.~\eqref{eq:S_Delta_h}.
Recall that the fractional particle shape is \emph{not} a simple scaling transform as the cases
considered thus far but is a \emph{shape transform}, which additionally changes the measure of its support.
The theory developed in Sec.~\ref{sec:stat_nonuniform} was for a scaling transform
only, keeping the particle shape unchanged. Therefore that theory cannot be used for 
quantitative comparison with the following simulations but can nevertheless serve as a
guideline to understanding the numerically observed behavior of $\Qmin$ and $\Hopt$. 

\begin{figure}[!h]
\begin{centering}
\includegraphics[width=0.7\textwidth]{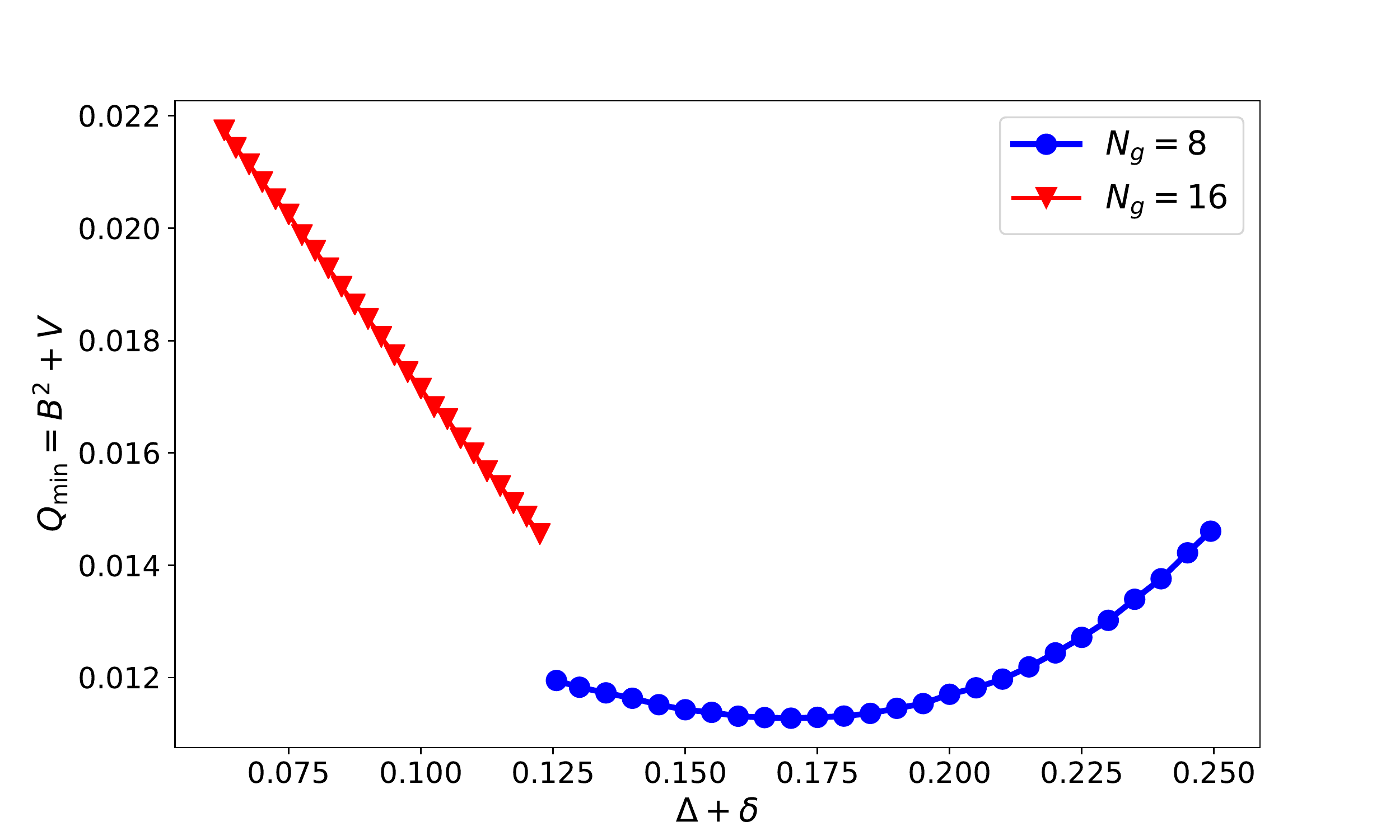} 
\par\end{centering}
\caption{Local bias-variance optimization at $x=1/2$ with
  the fractional width particle \eqref{eq:S_Delta_h}.
  The particle width $h=\Delta+\delta$ varies by changing $\delta$,
    but at the step discontinuity you have to change both $\delta$ and $\Delta$. (See the text.)}
\label{fig:BVO_fractional_particle} 
\end{figure}

These simulations were performed at $x=1/2$, for two \emph{fixed} grid sizes, $N_g=8$ and $16$, with
$N_p=1000$. The smaller number of particles was chosen to 
increase (for clarity) the relative numerical error due to the finite number of samples:
for $1000$ particles, the value of $\Qmin$ is expected to be about $10^{4/5}\approx 6.31$
times larger while the value of $\Hopt$ to be about $10^{1/5}\approx 1.58$ times larger.
As a guideline for the values of the minimal error and optimal width, we take the averages
of $\Qmin$ and $\Hopt$ between the two limiting shapes
($\Delta$-wide boxcar and $2\Delta$-wide linear tent);
these average values are $\Qmin \sim 0.0143$ and $\Hopt \sim 0.154$.
The simulation results in Fig.~\ref{fig:BVO_fractional_particle}
show minimum error $\Qmin\approx 0.0113$ and optimal width $\Hopt\approx 0.17$,
consistent with the above predictions. The observed discontinuity in the $Q(h)$
curve is also expected and easy to explain.
For the range $0.0625<h<0.125$ we use $\Delta=0.0625$ (or $N_g=16$),
and for $0.125<h<0.250$, we use $\Delta=0.125$ (or $N_g=8$).
The fractional width particle becomes a $\Delta$-wide boxcar shape for $\delta=0$
and a $2\Delta$-wide linear shape for $\delta=\Delta$. Therefore,
on the left side of the discontinuity the density is estimated by a linear particle shape
of width $0.125$ while on the right it is estimated by a boxcar shape of the same width.
Estimating the density by two different shapes (of the same width) is expected to produce different
errors $Q$ because of the different values of the shape coefficients $C_1$ and $C_2$
(see Table~\ref{table:Kernel_C_1_C_2}).

From the results in Fig.~\ref{fig:BVO_fractional_particle} we conclude that the
fractional width particle shape can indeed be used to attain
the BVO minimum error without having to change the type of particle
(or charge deposition rule).

\begin{table}[b!]
  \footnotesize
\centering
\begin{tabular}{c|c|c|c|c}
  \hline \rule{0pt}{3ex}
  \multirow{2}{*}{Shape}  & \multicolumn{2}{c|}{$\Qmin$} & \multicolumn{2}{c}{$\Hopt$}\\
  \cline{2-5}
 & theoretical  & numerical  & theoretical  & numerical \\
  \hline 
  \hline 
  Boxcar  & $0.00232$  & $0.00206$  & $0.0810$  & $0.0833$ \\
  \hline
  Quadr. spline  & $0.00223$  & $0.00198$  & $0.139$  & $0.136$ \\
  \hline 
  Trapezoidal  & $0.00219$  & $0.00194$  & $0.107$  & $0.107$ \\
  \hline 
  Epanechnikov  & $0.00219$  & $0.00194$  & $0.103$  & $0.107$ \\
  \hline \hline
\end{tabular}\caption{Comparison of theoretical values, Eqs.~\eqref{eq:h-opt} and \eqref{eq:Qmin},
  and numerical values of the optimal particle width and the error minimum. The more accurate
  value $\Qmin = 0.002186$ of the minimal error for Epanechnikov kernel is indeed slightly
  lower than the value of the trapezoidal shape; the table entry has been rounded off to three significant
  figures.}
\label{table:h0_Qmin} 
\end{table}

\section{Summary and conclusions}
\label{sec:conclusions}

We have presented analyses of the noise in particle methods used to
study electrostatic models in one dimension. We have described kernel
density estimation for continuous $x$, expressing the kernel in terms
of a fundamental kernel $K_f$ of width $1$, namely $K(x)=(1/h)K_{f}(x/h)$.
In this form its shape ($K_{f}$) and its width $h$ are represented
separately. Restricting our attention to uniform true electron density
$\rho(x)$ for these initial studies (and immobile ions of uniform, fixed density
$\rho_{i}(x)$ throughout the paper), we have computed the covariance matrix
$C(x,y)$ of the noise in the estimated electron density $\rho_{e}(x)$.
There are positive off-diagonal elements of the covariance matrix
related to the width $h$ of the kernel. But, more importantly, there
are constant negative elements, on and off the diagonal (the latter related to
negative correlations). These negative matrix elements arise from the fact that 
the total number of particles is fixed at
each time step. That is, for example, if the particles are concentrated
in one area (higher density estimate), they will be necessarily more
sparse (lower density estimate) in other areas. These negative correlations
lead to the property $\int\! C(x,y)dy=0$, i.e.~$C(x,y)$ has a zero
eigenvalue, $\int\! C(x,y)u(y)dy=0$ for eigenfunction $u(y)=1$.
We compute the estimated electric field from the estimated density by Gauss's law,
$\partial E/\partial x=\rho_{i}-\rho_{e}$, using $E(0)=E(1)$ by charge neutrality, 
$E(0)=0$ by periodicity. We also assume that the applied potential across the system
$\int_0^1 E(x)dx$ is zero. 
These boundary conditions and the negative correlations in $C(x,y)$ lead to properties of the noise
in the electric field related to a process called the
\emph{Ornstein-Uhlenbeck bridge}, described in Appendix A. The covariance matrix
of the electric field $C^{E}(x,y)$ is significantly reduced relative
to that of the commonly known Brownian process or the related Brownian
bridge, improving the fidelity of simulation results.
Because of the assumed
periodic boundary conditions on $[0,1]$, the covariance matrices
$C$ and $C^{E}$ have translational invariance properties, i.e.~$C(x,y)=C(x-y)$
and $C^{E}(x,y)=C^{E}(x-y)$. The latter also has an eigenfunction, namely $u(x)=1$,
with zero eigenvalue.

We have also investigated cases with non-constant density $\rho(x)$,
but still with continuous $x$. We have considered the total error
in the estimated density and analyzed it in terms of \emph{bias-variance
optimization} (BVO.) Small kernel widths have too few particles within
their support leading to too much variance;
for kernels with widths that are large compared to a
characteristic density gradient scale length,
the actual density is smoothed excessively.
The optimum between these two limits is found by BVO. The analysis also shows that
this optimum is weakly dependent on the kernel shape for kernels of equal widths.
We find that the scaling of the minimal error $\Qmin$ with the total number of particles $N_p$
is modified to $\Qmin\sim N_p^{-4/5}$ compared to the well known variance scaling $V\sim N_p^{-1}$.

We have analyzed these properties for a grid of discretized $x$
values. In this case the charge deposition rule is expressed in terms
of a \emph{particle shape}. We have discussed an important property
to be preserved in the discrete system: the exact preservation of
the net electron charge $\int\!\rho(x)dx$; the discrete version of 
this property is necessary
to ensure that the discretized electric field obeys the periodic boundary
conditions. If we assume that the particle shape obeys a sum rule
$\sum_{i}\Delta S(x_{i}-\xi)=\int\! S(x-\xi)\,dx=1$, saying that the
discretized integral over the shape equals the exact integral, then
exact preservation of charge holds. The particle obeys such a sum
rule, if it is the convolution of two kernels, $S(x)=\int\! K(y)\hat{K}(x-y)\,dy$,
where one of the kernels satisfies the sum rule.
An example of this occurs when one of the kernels is
a particle shape, $\hat{K}=S(x-y)$, or when finite elements
are used so that $\hat{K}(x-y)=\Psi(x-y)$ with $\Psi(x-y)$ having width equal
to an integral multiple of the grid spacing, $\Delta$,
and satisfying the sum rule.
The convolution formula appears naturally in variational particle
methods \cite{lewis:1970:136,evstatiev_variational_2013}.
It is important to note that the sum rule property for the particle
shape does not require that the kernel width is an exact multiple
of the grid spacing $\Delta$.

We have relaxed the approximations of the analytic
calculations of BVO optimization, doing numerical computations of
the total error as a function of the particle width.
The results show good agreement with the analytic results
over a range of particle shapes.

As practical applications of the results in this work, we have provided
evidence that noise correlations can be reduced by using sufficiently wide
particle shapes, decreasing finite number of particle numerical effects.
In non-uniform density, guidelines for the design, construction,
and implementation of computationally efficient particle shapes that
take advantage of the BVO is proposed. In particular, for large values of $\Hopt$
and small grid spacing $\Delta$ we recommend custom designed particles
of low (polynomial) order but sufficiently wide extent as a computationally
efficient alternative to the traditional spline shape functions. When
$\Hopt \sim \Delta$, we recommend fractional width particle shapes,
which can also be used to follow the optimal particle width in the
course of a simulation while keeping the charge deposition rule unchanged
and providing computational efficiency.

The bias-variance trade-off was discussed and shown to be important in the context of
PIC simulations in plasmas in Ref.~\cite{wu_qin_reducing_2018}.
In the present paper we stress the analytical development of this
idea as applied to particle-based numerical methods
\cite{finn_noise_2013,evstatiev_sherwood_2018,evstatiev_aps_2018}
and present detailed analysis of particle shapes related to
the width of the bias-variance minimum.
We also discuss exact charge conservation and the negative correlations
due to a fixed number of particles and their influence on the statistical
properties of the electric field.

\section*{Acknowledgments}

Sandia National Laboratories is a multimission laboratory managed and operated by
National Technology \& Engineering Solutions of Sandia, LLC, a wholly owned subsidiary of
Honeywell International Inc., for the U.S. Department of Energy's
National Nuclear Security Administration under contract DE-NA0003525.
This paper reviews objective technical results and analysis.
Any subjective views or opinions that might be expressed do not necessarily represent the views of
the U.S. Department of Energy or the U.S. Government.

The work of EGE was supported in part by NASA WV EPSCoR Grant \#NNX15AK74A and
in part by Sandia National Laboratory's LDRD Project \#209240.
The work of BAS was supported in part by the National Science Foundation Grant PHY-1535678.
The early work of JMF was supported by the University of California
UCOP program at Los Alamos National Laboratory. This manuscript has been assigned
Sandia No. \#SAND2021-1002 O

\begin{appendices}

\section{Appendix: Brownian bridge and Ornstein-Uhlenbeck bridge}
\label{app:brownian_bridge}

Consider a random walk on $0<t<T$ (continuous time),
with 
\begin{equation}
\frac{dv_{w}}{dt}=r_{w}(t),\label{eq:Wiener}
\end{equation}
where $v_{w}$ is the velocity of a Brownian particle and $r_{w}$
is its random acceleration, with $\langle r_{w}(t)r_{w}(s)\rangle=V_{0}\tau_{c}\delta(t-s)$.
Here, $V_{0}$ is the variance of $r_{w}(t)$ and $\tau_{c}$ is the
correlation time. We start by taking $v_{w}(0)=0$, and find $v_{w}(t)=\int_{0}^{t}r_{w}(\tau)d\tau$,
leading to 
\begin{equation}
  \langle v_{w}(t)v_{w}(s)\rangle
  =\int_{0}^{t}d\tau\int_{0}^{s}d\sigma\langle r_{w}(\tau)r_{w}(\sigma)\rangle=V_{0}\tau_{c}\text{min}(t,s),
  \label{eq:Wiener-Correlation}
\end{equation}
the standard result \cite{revuz_continuous_1999}.
In the analogy with the results of Sec.~\ref{sec:Electric-field-computation},
time takes the place of the distance $x$, the acceleration takes
the place of the density, the velocity takes the place of the electric
field, and the displacement takes the place of the electrostatic potential.

Now for each realization of the noise $r_{w}(t)$, consider the modified
process 
\begin{equation}
v_{b}(t)=v_{w}(t)-\frac{tv_{w}(T)}{T}.\label{eq:BrBrVariable}
\end{equation}
At this stage we still have $v_{b}(0)=0$. The usual random walk
has $\langle v_{w}(T)\rangle=0$ but this modified process has $v_{b}(T)=0$
for \emph{each} realization of the noise. This is the analog of the
condition $E(1)=E(0)$ of Sec.~\ref{sec:Electric-field-computation}. We
find 
\begin{equation}
\frac{dv_{b}}{dt}=r(t)=r_{w}(t)-\frac{v_{w}(T)}{T}\label{eq:Accel-BrBr}
\end{equation}
and $v_{b}(0)=0$, with the result 
\begin{equation}
C(t,s)\equiv\langle r(t)r(s)\rangle=V_{0}\tau_{c}\left[\delta(t-s)-\frac{1}{T}\right].\label{eq:Cov-mx-BB}
\end{equation}
This is proportional to the covariance in Eq.~\eqref{eq:delta-khat}. Notice
the stationarity condition $~C(t,s)=C(t-s)$, the analog of the translation
invariance condition in the spatial context, and $\int_{0}^{T}C(t-s)ds=0$,
as in Eq.~\ref{eq:int_C^E_zero}. We also find the covariance matrix
for the the Brownian bridge velocity $v_{b}(t)$,
\begin{equation}
  \langle v_{b}(t)v_{b}(s)\rangle=\int_{0}^{t}d\tau\int_{0}^{s}C(\tau,\sigma)d\sigma
  =V_{0}\tau_{c}\left[\text{min}(t,s)-ts\right].\label{eq:Br-Br-Cov-of-v}
\end{equation}
It is clear that the process of subtracting $tv_{w}(T)/T$ in Eq.~(\ref{eq:BrBrVariable})
is equivalent to integration of $dv_{b}/dt=r(t)$ with the covariance
matrix $r(t)$ given in Eq.~(\eqref{eq:Cov-mx-BB}).

The final step is to consider the \emph{Ornstein-Uhlenbeck bridge} \cite{Brownian_2nd_order2}
by starting with the system $dv/dt=r(t)$ as in Eq.~\eqref{eq:Accel-BrBr},
for now just relaxing the requirement $v(0)=0$. We find $v(t)=v_{0}+v_{1}(t)=v_{0}+\int_{0}^{t}r(\tau)d\tau$,
leading to 
\begin{equation}
\langle v(t)v(s)\rangle=C_{00}+C_{10}(t)+C_{01}(s)+C_{11}(t,s),\label{eq:velo-cov-OEBridge}
\end{equation}
where 
\[
  C_{00}=\langle v_{0}^{2}\rangle,\,\,C_{10}(x)=\langle v_{0}v_{1}(t)\rangle,\,\,C_{01}(y)
  =\langle v_{1}(s)v_{0}\rangle,\,\,\,\text{and}
\]
\[
C_{11}(t,s)=\int_{0}^{t}d\tau\int_{0}^{s}d\sigma\langle v_{1}(\tau)v_{1}(\sigma)\rangle,
\]
the exact analog of Eqs.~\eqref{eq:C00}-\eqref{eq:C-Electric}. The
net displacement $x(t)$ of the particle is found by integrating $dx/dt=v(t)$,
with $x(0)=0$, so that $x(t)=\int_{0}^{t}v(\tau)d\tau$. The Ornstein-Uhlenbeck
bridge modification is this: for each random walk, we choose $v_{0}$
so that the net displacement at $t=T$ is zero, $\int_{0}^{T}v(\tau)d\tau=0$.
This zero net displacement condition is the analog of the zero potential
difference requirement of Eq.~\eqref{eq:ZeroPotential}). The covariance
matrix $\langle v(t)v(s)\rangle$ is obtained by the methods outlined
in Sec.~\ref{sec:Electric-field-computation} and in Appendix~\ref{app:electric_field_covariance},
and illustrated in Fig.~\ref{fig:BB-plots}. Finally, the Ornstein-Uhlenbeck bridge
for smooth correlations,
$\langle r_{w}(t)r_{w}(s)\rangle=V_{0}\tau_{c}\delta(t-s)\rightarrow V_{0}\tau_{c}K(t-s)$
is treated in Appendix~\ref{app:electric_field_covariance} and in Fig.~\ref{fig:BB-plots}.

It is interesting to note that in analogy with Eq.~\eqref{eq:BrBrVariable},
we can relate the Ornstein-Uhlenbeck displacement variable $x(t)$
to the displacement for the Brownian bridge variable $x_{b}(t)$ having
$v_{0}=0$ by 
\[
x(t)=x_{b}(t)-\frac{tx_{b}(T)}{T}.
\]
The Brownian Bridge defined above has the requirement that the particle
velocity return to zero at $t=T$. The Ornstein-Uhlenbeck bridge has
the further requirement that the particle \emph{displacement} return to its
original position, i.e.~that the average velocity $\int_{0}^{T}v(t)dt/T$
be zero.

\section{Appendix: Electric field covariance matrix for general kernels}
\label{app:electric_field_covariance}

In this appendix we derive the covariance matrix for the electric
field from Eq.~\eqref{eq:C-E-initial-expression}, relaxing the special case of
Eq.~\eqref{eq:delta-khat} to a general kernel,
\begin{equation}
C(x,y)=K_{0}(x-y)-1,\label{eq:GeneralKernel}
\end{equation}
where the $1/N_{p}$ factor has been suppressed and $K_{0}(x)=(1/h)K_{f}(x/h)$.
We take $K_{0}(x)\rightarrow K(x)$ to be extended to be periodic
of period $1$, so that it is even about $x=1/2$. As in Eq.~\eqref{eq:C-E-initial-expression},
we conclude 
\begin{equation}
C^{E}(x,y)=C_{00}+C_{10}(x)+C_{01}(y)+C_{11}(x,y).\label{eq:C^E-four-terms}
\end{equation}

For simplicity we pick the fundamental kernel $K_{f}(x)$ to be the boxcar. 
We find $XC^{E}=(\partial_{x}+\partial_{y})C^{E}=0$, which implies
translational invariance $C^{E}(x,y)=C^{E}(x-y)$, and symmetry $C^{E}(x,y)=C^{E}(y,x)$ implies
\[
C^{E}(x,y)=C^{E}(|x-y|).
\]
Finally, these relations imply that $C^{E}(x,y)$ is periodic in $x-y$
with period $1$. 

An alternate approach begins with
$C(x,y)=\langle\tilde{\rho}(x)\tilde{\rho}(y)\rangle=\langle\tilde{E}'(x)\tilde{E}'(y)\rangle$.
This leads to
\begin{equation}
  C(x,y)=\frac{\partial^{2}}{\partial x\partial y}C^{E}(x,y)
  =\frac{\partial^{2}}{\partial x\partial y}C^{E}(x-y),
  \label{eq:PDE-for-C^E}
\end{equation}
with the last step following from translational invariance.
Eq.~\eqref{eq:PDE-for-C^E} leads to:
\begin{equation}
C(x-y)=-\partial_{x}^{2}C^{E}(x-y).\label{eq:PDE-for-C^E-2}
\end{equation}
For $C(x)=\delta(x)-1$ (ignoring $1/N_{ppc}$ factor), we find $\partial_{x}^{2}C^E(x)=-C(x)=-\delta(x)+1$
implies
\begin{equation}
  C^{E}(x)=D_{0}-\frac{1}{2}|x|+\frac{x^{2}}{2}.
  \label{eq:C^E-delta-function-example}
\end{equation}
This satisfies $C^{E}(1)=C^{E}(0)=D_{0}$ or $C^{E}(-1/2)=C^{E}(1/2)$,
related to the boundary condition on the electric field from overall
charge neutrality. The condition $\int_{0}^{1}C^{E}(x)dx=0$, from
the zero applied potential condition $\int_{0}^{1}E(x)dx=0$, leads
to $D_{0}=1/12$. These results are in agreement with Eqs.~\eqref{eq:Full-CE(x,y)} 
and \eqref{eq:CE_translation_invariant} when the factor $1/N_p$ is reinstated.

Finally, for the linear kernel, $C(x)=K_{fL}(x/h)/h-1$,
we solve $\partial_{x}^{2}C^{E}(x)=-K_{fL}(x/h)/h+1$.
The neutrality requirement $C^E(-1/2)=C^E(1/2)$ is easily seen to be
satisfied. These results
as well as those from $h=0$ (Eqs.~\eqref{eq:C^E-delta-function-example},
\eqref{eq:Full-CE(x,y)}, \eqref{eq:CE_translation_invariant})
are plotted in Fig.~\ref{fig:BB-plots}.
Note that the cusp at $x=y$ for $h=0$ is smoothed
and relation $\int_{0}^{1}C^{E}(x,y)dy=0$ is found to hold.
With the $1/N_p$ factor, these results agree with the results derived by the method above (see
Eq.~(\ref{eq:C^E-four-terms})).

\section{Appendix: Scaling of the kernel}
\label{app:kernel_scaling}

The information specific to a given kernel is contained in the shape
coefficients $C_1$ and $C_2$ [cf. Eq.~\eqref{eq:C_1_C_2}].
Sometimes it may be more convenient to work with the scaled kernel \eqref{eq:K_definition}
instead of the fundamental kernel.
Additionally, published literature may define the fundamental
kernel with a width different from unity.
To make a connection between scaled or differently defined kernels
and the fundamental kernel as defined in our presentation, we examine how
the coefficients $C_1$ and $C_2$ scale under a scaling transformation
$h\rightarrow \alpha h$ for arbitrary $h$ and scaling factor $\alpha>0$.
For example, to obtain the linear particle shape in Table~\ref{table:particle_shapes}
from the linear fundamental kernel in Table~\ref{table:fundamental_kernels},
we use Eq.~\eqref{eq:K_definition} with $h=\alpha\Delta$ with $\alpha=2$;
similarly, for the quadratic spline and trapezoidal particles we use $\alpha=3$, etc.
(We remind the reader that not all fundamental kernels allow for a scaling transform
leading to a particle shape satisfying the sum rule, e.g., the Epanechnikov fundamental kernel.)

The kernel scales as 
\[
  K(x)=\frac{1}{h}K_{f}\left(\frac{x}{h}\right)\rightarrow\frac{1}{\alpha h}K_{f}\left(\frac{x}{\alpha h}\right)
  =\frac{1}{\alpha}K_{f}\left(\frac{\zeta}{\alpha}\right),
\]
where $\zeta=x/h$. It is easy to verify that the scaled kernel is also normalized to
unity [cf. Eq.~\eqref{eq:S_normalization-1}].
Now we calculate the scaled coefficients, $C_{s,1}$ and $C_{s,2}$. We have
\[
  C_{s,1} 
  = \int\!\! d\zeta\, \frac{1}{\alpha^{2}}K_{f}^{2}\left(\frac{\zeta}{\alpha}\right)
  =\frac{1}{\alpha}\int\!\! d\gamma\, K_{f}^{2}(\gamma)=\frac{1}{\alpha}C_{1},
\]
\[
  C_{s,2} 
  = \int\!\! d\zeta\, \zeta^{2}\frac{1}{\alpha}K_{f}\left(\frac{\zeta}{\alpha}\right)
  =\alpha^{2}\int\!\! d\gamma\, \gamma^{2}K_{f}(\gamma)=\alpha^{2}C_{2},
\]
with $\gamma=\zeta/\alpha$. Again, formulas \eqref{eq:C_1_C_2}
are based on the fundamental kernel and therefore yield the values of
$C_{1}$ and $C_{2}$ on the right hand sides above, as seen in Table~\ref{table:Kernel_C_1_C_2}.
Calculating the scaled values of $\Qmin$, $\Hopt$, and $W_{\rm Q}$ we get 
\[
  Q_{{\rm s,min}}\sim\left(C_{s,1}\sqrt{C_{s,2}}\right)^{4/5}
  =\left(\frac{C_{1}}{\alpha}\sqrt{\alpha^{2}C_{2}}\right)^{4/5}
  =\left(C_{1}\sqrt{C_{2}}\right)^{4/5}\sim \Qmin\,,
\]
\[
  h_{{\rm s,opt}}\sim W_{\rm s,Q} \sim
  \left(\frac{C_{s,1}}{C_{s,2}^{2}}\right)^{1/5}
  =\left(\frac{1}{\alpha}C_{1}\frac{1}{\alpha^{4}C_{2}^{2}}\right)^{1/5}
  =\frac{1}{\alpha}\left(\frac{C_{1}}{C_{2}^{2}}\right)^{1/5}
  \sim \frac{1}{\alpha}\,\Hopt \sim \frac{1}{\alpha} W_{\rm Q}.
\]
We see that $\Qmin$ remains {\emph{unchanged}}, while $\Hopt$ and $W_{\rm Q}$ \emph{scale inversely}
with the scaling factor $\alpha$.

\end{appendices}

\bibliography{PIC_Noise_manuscript}
\bibliographystyle{unsrt}

\end{document}